\documentclass[a4paper,fleqn,usenatbib]{mnras}

\usepackage{amsmath}	
\usepackage{amssymb}	
\usepackage{newtxtt,newtxmath}

\usepackage[T1]{fontenc}
\usepackage{ae,aecompl}

\usepackage{graphicx}	
\usepackage{epstopdf}
\usepackage{subcaption}
\usepackage{multicol}
\usepackage{multirow}
\usepackage{float}
\usepackage[usenames,dvipsnames,svgnames,table]{xcolor}
\usepackage{enumitem}
\usepackage{bm}
\usepackage{tablefootnote}

\usepackage[noabbrev]{cleveref}

\crefname{equation}{}{}
\crefname{figure}{}{}

\newcommand{\lya}{Ly$\alpha$}
\newcommand{\ha}{H$\alpha$}

\newcommand{\oiii}{[O{\sc iii}]}
\newcommand{\oii}{[O{\sc ii}]}

\newcommand{\msol}{M$_\odot$}
\newcommand{\wtheta}{$w(\theta)$}
\newcommand{\ro}{$r_0$}
\newcommand{\muv}{M$_\textrm{UV}$}

\title[Clustering Properties of LAEs~at $z\sim2 - 6$]{The clustering of typical \lya~emitters from $\bm{z \sim 2.5 - 6}$: host halo masses depend on \lya~and UV luminosities}
\author[Khostovan et al.]{A.~A.~Khostovan$^{1}$\thanks{NASA Earth and Space Science Fellow}\thanks{E-mail:
akhostov@gmail.com}, D.~Sobral$^{2,3}$, B.~Mobasher$^{1}$, J.~Matthee$^{3,4}$\thanks{Zwicky Fellow}, R.~K.~Cochrane$^{5}$, \newauthor 
N.~Chartab~Soltani$^{1}$, M.~Jafariyazani$^{1}$, A.~Paulino-Afonso$^{2,6,7}$, S.~Santos$^{2}$, J.~Calhau$^{2}$ \\
$^{1}$Department of Physics \& Astronomy, University of California, Riverside, United States of America\\
$^{2}$Department of Physics, Lancaster University, Lancaster, LA1 4YB, UK \\
$^{3}$Leiden Observatory, Leiden University, PO Box 9513, NL-2300 RA Leiden, the Netherlands\\
$^{4}$Department of Physics, ETH Z\"{u}rich,Wolfgang - Pauli - Strasse 27, 8093 Z\"{u}rich, Switzerland\\
$^{5}$SUPA, Institute for Astronomy, Royal Observatory of Edinburgh, Blackford Hill, Edinburgh EH9 3HJ, UK\\
$^{6}$Instituto de Astrof\'isica e Ci\^encias do Espa\c{c}o, Universidade de Lisboa, OAL, Tapada da Ajuda, P-1349-018 Lisboa, Portugal\\
$^{7}$Departamento de F\'isica, Faculdade de Ci\^encias, Universidade de Lisboa, Edif\'icio C8, Campo Grande, P-1749-016 Lisboa, Portugal}

\date{}

\pubyear{2018}
\begin{document}

\label{firstpage}
\pagerange{\pageref{firstpage}--\pageref{lastpage}}
\maketitle

\begin{abstract} 
We investigate the clustering and halo properties of $\sim 5000$ \lya-selected emission line galaxies (LAEs) from the Slicing COSMOS 4K (SC4K) and from archival NB497 imaging of SA22 split in 15 discrete redshift slices between $z \sim 2.5 - 6$. We measure clustering lengths of $\textrm{\ro} \sim 3 - 6\ h^{-1}$ Mpc and typical halo masses of $\sim 10^{11}$ \msol~for our narrowband-selected LAEs with typical $L_\textrm{\lya} \sim 10^{42 - 43}$ erg s$^{-1}$. The intermediate band-selected LAEs are observed to have $\textrm{\ro} \sim 3.5 - 15\ h^{-1}$ Mpc with typical halo masses of $\sim 10^{11 - 12}$ \msol~and typical $L_\textrm{\lya} \sim 10^{43 - 43.6}$ erg s$^{-1}$. We find a strong, redshift-independent correlation between halo mass and \lya~luminosity normalized by the characteristic \lya~luminosity, $L^\star(z)$. The faintest LAEs ($L \sim 0.1\ L^\star(z)$) typically identified by deep narrowband surveys are found in $10^{10}$ \msol~halos and the brightest LAEs ($L \sim 7\ L^\star(z)$) are found in $\sim 5 \times 10^{12}$ \msol~halos. A dependency on the rest-frame 1500 \AA~UV luminosity, \muv, is also observed where the halo masses increase from $10^{11}$ \msol~to $10^{13}$ \msol~for \muv$\sim -19$ mag to $-23.5$ mag. Halo mass is also observed to increase from $10^{9.8}$ \msol~to $10^{12.3}$ \msol~for dust-corrected UV star formation rates from $\sim 0.6$ \msol~yr$^{-1}$ to $10$ \msol~yr$^{-1}$ and continues to increase up to $10^{13.5}$ \msol~in halo mass, where the majority of those sources are AGN. All the trends we observe are found to be redshift-independent. Our results reveal that LAEs are the likely progenitors of a wide range of galaxies depending on their luminosity, from dwarf-like, to Milky Way-type, to bright cluster galaxies. LAEs therefore provide unique insight into the early formation and evolution of the galaxies we observe in the local Universe.

\end{abstract}


\section{Introduction}
The current state of galaxy formation and evolution theory suggests that galaxies formed with the assistance of their host dark matter halos, where deep gravitational potential wells allowed for the accretion of cold gas to form the galaxies and fuel star formation activity (see \citealt{Benson2010} and \citealt{Somerville2015} for a review). The era between cosmic noon (peak of cosmic star-formation activity; $z \sim 2$) and the `end' of the epoch of reionization ($z \sim 6$) constitutes an important time period in the history of the Universe. It is within that time ($\sim 2$ Gyr) that galaxies rapidly evolved with their typical star formation rates increasing by an order of magnitude (e.g., \citealt{Madau2014,Khostovan2015}). Since galaxies reside and evolve within dark matter halos, the host halos likely play a fundamental role in the overall evolution of galaxies. How do galaxies and their host halos co-evolve? 

Addressing such a fundamental question requires large samples of high-$z$ galaxies with well understood selection functions. Since galaxies reside in dark matter halos, their spatial clustering directly traces the host dark matter halos, although with a few assumptions (e.g., halo mass and bias functions, occupation distributions; see \citealt{Cooray2002} for a review).

One class of high-$z$ star forming galaxies that can be used to investigate clustering properties are \lya~emitters (LAEs). These are typically young, low-mass (e.g., \citealt{Gawiser2006,Finkelstein2007,Guaita2011,Hagen2016,Shimakawa2017,Hao2018,Kusakabe2018b}), compact galaxies (e.g., \citealt{Malhotra2012,Kobayashi2016,Paulino2018}) with high rest-frame equivalent widths (e.g., \citealt{Malhotra2002,Hu2010,Ciardullo2012,Zheng2014,Hashimoto2017}), low metallicities, and high ionization states (e.g., \citealt{Finkelstein2011,Erb2016,Trainor2016}).

Samples of \lya~emitters (LAEs) selected via narrowband surveys provide an efficient and robust window for probing the high-$z$ Universe (e.g., \citealt{Rhoads2000,Ouchi2008,Nilsson2009}). Narrowband surveys have the added advantage of enabling large samples of galaxies by directly observing emission lines associated with star formation and AGN activity using specially designed photometric filters. Because the filter widths are quite narrow (between $50 - 200$ \AA~in FWHM), emission line galaxies selected in narrowband surveys have reliable redshifts within $1 - 2$ percent error (albeit with typical $5 - 10$ percent contamination; see e.g., \citealt{Sobral2018}) and are prime sources for future spectroscopic follow-up studies.

Previous narrowband surveys have searched for LAEs at $z \sim 2 - 7$ (e.g., \citealt{Cowie1998,Rhoads2000,Gronwall2007,Ouchi2008,Matthee2015,Santos2016,Konno2018,Sobral2018}), but only a few studies have investigated their clustering properties. The earliest work on LAE clustering was done by \citet{Ouchi2003}, which observed 87 LAEs at $z = 4.86$ in the 0.15 deg$^2$ Subaru Deep Field and reported the first angular correlation functions and clustering lengths for LAEs ($\textrm{\ro} = 3.5\pm0.3\ h^{-1}$ Mpc). Subsequent narrowband surveys have allowed for measurements of LAE clustering properties at $z\sim 2 - 7$ (e.g., \citealt{Shimasaku2004,Kovac2007,Ouchi2010,Zheng2016,Hao2018,Ouchi2018}). \citet{Guaita2010} presented the first $z \sim 2.1$ measurement using a sample of 250 LAEs in the $0.36$ deg$^2$ ECDF-S field and found $\textrm{\ro} = 4.8\pm0.9$ Mpc $h^{-1}$ and a typical halo mass of $\sim 3.2 \times 10^{11}$ \msol. The recent $z \sim 2$ measurements of \citet{Kusakabe2018} presented the latest constraints using a sample of $\sim 1250$ LAEs in four separate fields for a total survey area of 1 deg$^2$ and found $\textrm{\ro} = 2.38^{+0.34}_{-0.39}$ Mpc $h^{-1}$ and a typical halo mass of $4 \times 10^{10}$ \msol. Both surveys cover the same redshift, but report significantly different results. In respect to \citet{Guaita2010}, the survey area of \citet{Kusakabe2018} is about 3 times larger and 2 times fainter in line flux. Given that the \lya~luminosity functions are steep ($\alpha ~ -1.8$; e.g., \citealt{Konno2016,Sobral2018}), deeper samples will be dominated by faint LAEs, which highlights the importance of investigating the clustering properties in terms of their \lya~luminosities.

Although the past two decades have produced a handful of LAE clustering measurements, only a few have focused on how the measured clustering properties are correlated to the physical properties of LAEs (e.g., line luminosity, stellar mass, star formation rates). \citet{Ouchi2003} showed that the brightest $z \sim 4.8$ LAEs are more clustered than the faint LAEs. \citet{Bielby2016} presented clustering measurements of $z\sim3.1$ LAEs in bins of $R$-band limiting magnitude (corresponding to the 1500\AA~UV continuum luminosity, \muv) and found $\textrm{\ro} \sim 3\ h^{-1}$ Mpc at $R_{lim} = 27.5$ mag and $\textrm{\ro} \sim 4.5\ h^{-1}$ Mpc at $R_{lim} = 26$ mag, such that LAEs with high UV continuum luminosities (a tracer of star formation) are more strongly clustered. Recently, \citet{Kusakabe2018} found a weak correlation between \lya~line luminosity limit and clustering length/halo mass. This suggests that the \lya~and UV continuum luminosity, which to first order trace star formation activity, correlate with halo mass. However, sample selection effects and varying survey depths prohibit detailed analysis of the evolution and origin of such trends.

Besides \lya~studies, narrowband surveys focused on \ha~\citep{Sobral2010,Cochrane2017,Cochrane2018}, \oiii~\citep{Khostovan2018}, and \oii~\citep{Khostovan2018} emission line-selected galaxies have also found strong trends between the physical properties of star-forming/active galaxies and their host halo properties. These surveys reveal strong correlations between halo mass and line luminosity (proxy for star formation rate) up to $z\sim2.23$ (\ha), $z\sim3.3$ (\oiii), and $z\sim4.7$ (\oii). \citet{Khostovan2018} found that these trends are redshift-independent once the evolution in typical line luminosity is taken into account suggesting that the host halo and residing galaxy co-evolve in unison over cosmic time.  

Although much work has been done on quantifying the clustering/halo properties and its relation to the physical properties of star-forming galaxies, not much focus has been applied to such analysis with \lya-selected samples, which allow for observing such trends up to the epoch of reionization. In this paper, we use the Slicing COSMOS 4K (SC4K) survey and archival NB497 imaging to investigate the clustering properties of LAEs in 15 discrete redshift slices between $z \sim 2.5 - 5.8$ with a total of $\sim 5000$ LAEs within the 2 deg$^2$ COSMOS and $1.3$ deg$^2$ SA22 fields, corresponding to comoving volumes between $1 - 6 \times 10^6$ Mpc$^3$ and total comoving volume of $\sim 6 \times 10^7$ Mpc$^3$ for the full survey.

The paper is organized as follows: In \S \ref{sec:sample}, we describe the sample of LAEs. In \S \ref{sec:clustering_measurements}, we present how we generate our random samples, the methodology in measuring the angular correlation functions, the clustering length, and halo masses, as well as corrections for cosmic variance and a discussion on contamination. In \S \ref{sec:UV_prop}, we show our methodology in measuring the UV continuum luminosity and slope, as well as the UV star formation rates. In \S \ref{sec:results}, we present our main results with discussion regarding its interpretation. We present our main conclusions and final remarks in \S \ref{sec:conclusion}.

Throughout this paper we assume a $\Lambda$CDM cosmology with H$_0 = 70$ km s$^{-1}$ Mpc$^{-1}$, $\Omega_m = 0.3$, and $\Omega_\Lambda = 0.7$. Our reported values for halo masses and \ro~are in units of \msol~$h^{-1}$ and Mpc $h^{-1}$, respectively, unless otherwise specified. Our measurements of star formation rates assume a Salpeter IMF.

\section{\lya~Sample}
\label{sec:sample}

\subsection{Slicing COSMOS 4K at \boldmath$2.5 < z < 6$}
\label{sec:SC4K}

Our sample is drawn from the publicly available Slicing COSMOS in 4K survey (SC4K; \citealt{Sobral2018,Paulino2018}), which contains 3908 \lya~emitters (LAEs). The survey uses Subaru imaging from 12 intermediate bands \citep{Capak2007,Taniguchi2007,Taniguchi2015} in the $\sim 2$ deg$^{2}$ COSMOS field \citep{Scoville2007,Capak2007} that are reanalyzed following the procedures outlined in \citet{Sobral2018}. The SC4K survey also includes imaging using four narrowband filters: NB392 ($z = 2.2$; \citealt{Sobral2017_CALYMHA, Matthee2017}), NB501 ($z = 3.1$; \citealt{Matthee2017}), NB711 ($z = 4.8$; \citealt{Sobral2018}), and NB816 ($z = 5.7$; \citealt{Santos2016}). The NB392 and NB501 observations were conducted using the Wide Field Camera on the Isaac Newton Telescope, while all other narrowband and intermediate band observations are from archival Subaru imaging. We restrict our analysis to the samples with $z \gtrsim 2.5$ and also those samples for which the image-to-image variation is negligible and spatially contiguous\footnotemark. This includes all 12 intermediate bands and the NB711 and NB816 narrowband samples. 
 
\footnotetext{Although SC4K has a total of 4 narrowband filters, we restrict ourselves to the NB711 and NB816 samples. This is because the NB392 and NB501 images are not spatially contiguous (i.e., chip gaps), which requires special care when generating the random/mock samples (see \S\ref{sec:random_sample}). The other issue with the NB392 and NB501 images is the inhomogeneous depths. Image-to-image depth variation has to also be carefully taken into account to properly generate the random samples. Because, of these limitations, we decided to exclude them from the final sample of LAEs used in this study.}

We refer the reader to \citet{Sobral2018} for details regarding the sample selection. In brief, initial emission line galaxy candidates were selected by applying a rest-frame equivalent width cut of $25$ \AA~and $50$ \AA~for narrowbands and intermediate bands, respectively, along with a nebular excess significance cut of $\Sigma > 3$. A combination of spectroscopic redshifts, photometric redshifts, and color-color diagnostics were used to select potential LAEs. These candidates were then visually checked to remove any contaminants arising from artifacts not removed in the catalog generation (e.g., diffraction patterns, edge effects resulting in poor S/N) and sources that have their narrow or intermediate band photometry boosted by the presence of a bright halo from a nearby star in the image. In total, a final sample size of 3702 LAEs spanning between $z \sim 2.5 - 6$ were selected based on the 12 intermediate bands and 2 narrowband filters used in this study. In total, 102 of the 3702 LAEs have spectroscopic confirmation.

Table \ref{table:muv_filters} highlights the redshifts and sample sizes of all the LAE redshift slices. To take advantage of larger sample sizes, especially at the high-$z$ end (e.g. IA767 and IA827), we combine the intermediate band samples to form five larger samples in redshift bins as described at the bottom of Table \ref{table:muv_filters}. The choice of combinations was based on maximizing the sample size, with the different completeness limits per individual sample taken into account. This also includes minimizing the redshift widths of the final combined sample so as to remove possible cosmic evolutionary effects when using the samples in our clustering measurements.

\subsection{SA22 NB497 at \boldmath$z = 3.1$}
In addition to the SC4K sample, we also use a sample of 1198 $z = 3.1$ LAEs observed in the $1.38$ deg$^2$ SA22 field using archival NB497 imaging \citep{Matsuda2004,Yamada2012}. The observations were done using Suprime-Cam on the Subaru 8.2 m telescope and consisted of 7 contiguous, homogeneous pointings, which also covered the large SSA22 protocluster identified by \citet{Steidel1998}. The data and source selection was done independently of \citet{Yamada2012} and followed the methodology of \citet{Matthee2017}. CFHT MegaCam $ugi$ photometry from the CFHTLS survey was used in the source selection as opposed to the original Subaru SuprimeCam $BV$ photometry used in \citet{Yamada2012}. All LAE candidates were selected with a rest-frame equivalent width cut of $25$ \AA~and with a similar color-color selection criteria used for $z \sim 3$ LBG candidates \citep{Hildebrandt2009}. Details regarding the source selection are presented in an upcoming paper (Matthee et al., in prep). Of the 1198 $z = 3.1$ LAEs, 54 of them are spectroscopically confirmed from observations by \citet{Yamada2012} and \citet{Saez2015}.

\section{Clustering Measurements}
\label{sec:clustering_measurements}

\begin{figure}
	\centering
	\includegraphics[width=\columnwidth,trim=4 4 4 4,clip=True]{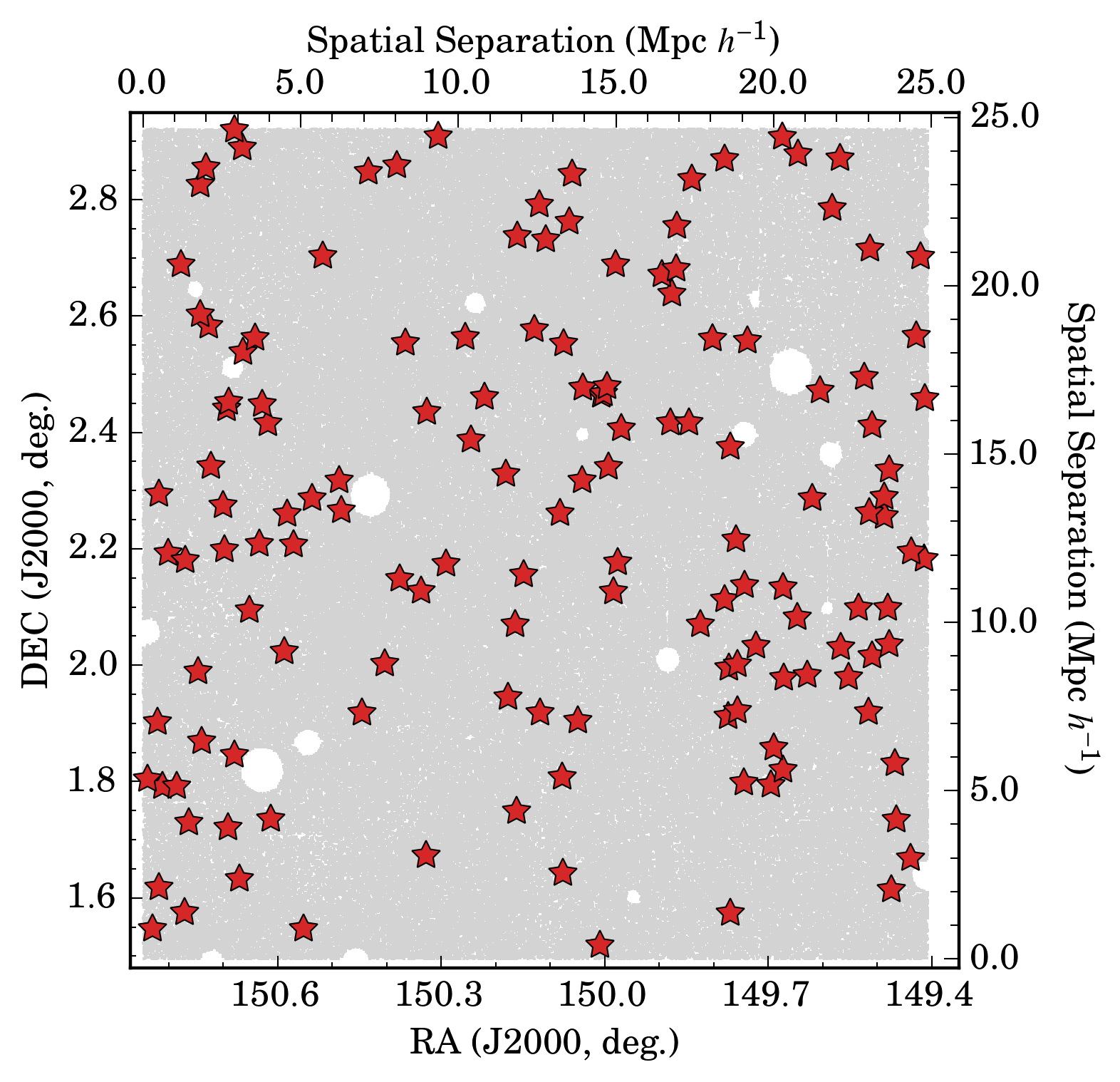}
	\caption{An example of the on-sky distribution using the $z = 4.13$ IA624 sample of LAEs shown as red stars. Specific regions are masked throughout the field (shown as empty regions within the grey shaded areas) to account for bright stars and various artifacts. These masked regions are taken into account when generating the random catalogs. The flux depths throughout the images are homogeneous which decreases the effects of image-to-image variations in the clustering measurements.}
	\label{fig:masking}
\end{figure}

\subsection{Random Sample}
\label{sec:random_sample}

We adopt the approach used in \citet{Khostovan2018}, which followed \citet{Sobral2010} to generate random samples. For each narrowband and intermediate band sample, we use the corresponding masked regions maps (see \citealt{Sobral2018}) to remove parts of the survey where the imaging is poor in quality or affected by diffraction patterns around bright stars. Figure \ref{fig:masking} shows an example of the masked regions overlaid with our $z = 4.13$ IA624 LAE sample. The bluer intermediate bands (IA427 - IA574) also include an extra $\sim 0.02$ deg$^2$ masking due to the smaller $u$-band imaging area, which is necessary to distinguish LAEs from other line emitters. Each sample is homogeneous in depth (see \citealt{Sobral2018} for details) which allows us to exclude the effects of variable depth in generating the mock random samples. In total, our random samples consist of $\sim 10^6$ mock sources per corresponding intermediate/narrowband sample, for which a subset is selected randomly when making measurements of the angular correlation functions (see \S\ref{sec:acf}).

\subsection{Angular Correlation Function}
\label{sec:acf}
The two-point correlation function is a statistical tool that traces the clustering properties of a given sample by comparing the angular (or spatial) distribution to a random distribution \citep{Peebles1980}. The angular two-point correlation function is typically defined as:
\begin{eqnarray}
\mathrm{d}P_{12} = \mathcal{N} (1 + w(\theta_{12})) \mathrm{d}\Omega_1 \mathrm{d}\Omega_2
\end{eqnarray}
where $\mathrm{d}P_{12}$ is the probability of finding two galaxies at positions $\Omega_1$ and $\Omega_2$ with an angular separation of $\theta_{12}$ for a complete sample with number density, $\mathcal{N}$. In the case that no angular/spatial correlation exists, then \wtheta$ = 0$ and the probability reduces into a uniform distribution. Therefore, a nonzero correlation function indicates that galaxies are clustered on the sky.
	
To measure the angular correlation function (ACF), we use the methodology proposed by \citet{Sobral2010} and modified by \citet{Khostovan2018}. This methodology uses the \citet[LS]{Landy1993} estimator to measure the observed correlation function described as:
\begin{eqnarray}
	\textrm{\wtheta} = 1 + \Bigg(\frac{N_R}{N_D}\Bigg)^{2} \frac{DD(\theta)}{RR(\theta)} - 2 \Bigg(\frac{N_R}{N_D}\Bigg) \frac{DR(\theta)}{RR(\theta)}
	\label{eqn:LS_estimator}
\end{eqnarray}
where $DD$, $DR$, and $RR$ are the number of data-data, data-random, and random-random galaxy pairs, respectively, and $N_R$ and $N_D$ are the number of random and data (real) galaxies, respectively. The errors are described as:
\begin{eqnarray}
	\Delta \textrm{\wtheta} = \frac{1+\textrm{\wtheta}}{\sqrt{DD(\theta)}}
	\label{eqn:errors}
\end{eqnarray}
and are assumed to be Poisson errors. \citet{Norberg2009} showed that Poisson statistics underestimates the ``true'' errors and suggests using bootstrapping for sample sizes of the order of $10^7$ sources. \citet{Khostovan2018} find no significant difference between the two error assessments for sample sizes similar to the one used in this study (see Appendix C of \citealt{Khostovan2018}).

We use the following fitting function:
\begin{equation}
	w(\theta) = A_w \Bigg(\theta^\beta - \frac{\sum{RR \theta^\beta}}{\sum{RR}}\Bigg)
	\label{eqn:wtheta_PL}
\end{equation}
where the second term is the integral constraint (IC), which takes into account the underestimation of \wtheta~at large angular separations and  $A_w$ and $\beta$ are the clustering amplitude and slope, respectively. Since our sample sizes are not large enough to constrain the slope, we assume $\beta = -0.8$ (fiducial; see \citealt{Peebles1980}) for all our measurements. 

As described in \citet{Khostovan2018}, we measure the correlation function by randomly selecting a starting bin center between $1'' - 5''$ (e.g. $8 - 40$ kpc at $z \sim 2.5$, $6 - 30$ kpc at $z \sim 5.7$), constant bin sizes of $\log_{10} \Delta\theta \sim 0.1'' - 0.3''$, and a maximum angular separation of $7200''$ (e.g., 58 Mpc at $z \sim 2.5$, 43 Mpc at $z \sim 5.7$). The random sample is drawn from our large mock catalog mentioned in \S\ref{sec:random_sample}. We randomly select the number of mock galaxies to be $10 - 500$ times the number of observed LAEs. With both the real and random samples, we measure \wtheta~via the LS estimator as described in Equation \ref{eqn:LS_estimator}. We then fit our power law model as described in Equation \ref{eqn:wtheta_PL} to measure the clustering amplitude. 

This whole process is iterated 2000 times while varying the bin sizes, centers, and random sample sizes per redshift slice. The reported clustering amplitude measurements and their associated errors are based on the median of the $A_w$ distribution drawn from all realizations with the errors being a combination of the scatter (random) in the distribution and also the median (systematic) error of all 2000 realizations added in quadrature. The benefit of this approach (see also \citealt{Sobral2010}), as highlighted by \citet{Khostovan2018}, is that it takes into account the systematic effects due to bin selection (e.g., centers, widths), which are especially important for small sample sizes ($< 100$ sources).

\subsection{Real Space Correlation Function}
\label{sec:spatial_corr}
The real space correlation function, $\xi(r)$, measures the spatial clustering of galaxies and is typically described as a power law of the form $\xi(r) = (r/r_0)^\gamma$, where $r_0$ is the clustering length and $\gamma$ is the slope of the correlation function. The angular correlation function (see \S\ref{sec:acf}) is a projection of the spatial correlation function and typical clustering studies relate the two using the Limber approximation \citep{Limber1953}. Although this works for typical redshift surveys, \citet{Simon2007} showed quantitatively that the approximation fails, especially at large angular separations and when the width of the redshift distributions become similar to a delta function\footnotemark. As a consequence, at large angular separations the slope of the angular correlation function changes from $\gamma+1$ to $\gamma$, such that \wtheta~is a rescaled version of $\xi(r)$. Various narrowband studies have observed this rescaling at a wide range of redshifts (e.g., \citealt{Gawiser2007,Guaita2010,Sobral2010,Cochrane2017,Khostovan2018,Kusakabe2018,Ouchi2018}).

\footnotetext{\citet{Simon2007} showed that using the Limber approximation, the clustering amplitude is related to the spatial clustering length as $A_w \propto r_0^{1-\gamma}/(2 \Delta r)$, with $\Delta r$ being the comoving filter width. As redshift distributions become narrower ($\Delta r \rightarrow 0$), the approximation breaks down and produces unphysical measurements for $A_w$.}

To properly measure the spatial correlation function for our samples requires that we use the exact form of the Limber equation. We follow the methodology of \citet{Khostovan2018} which uses the exact Limber equation as defined by \citet{Simon2007}:
\begin{align}
\centering
\textrm{\wtheta} & = \frac{r_0^{-\gamma}}{1+ \cos \theta} \int\limits_0^{\infty} \int\limits_{\bar{r}\sqrt{2(1-\cos\theta)}}^{2\bar{r}} \frac{2 p(\bar{r} - \Delta) p(\bar{r} + \Delta)]}{R^{-\gamma - 1} \Delta} \mathrm{d}R \mathrm{d}\bar{r} \nonumber \\
\Delta & =  \sqrt{\frac{R^2 - 2\bar{r}^2 (1-\cos{\theta})}{2 (1+\cos{\theta})}}
\label{eqn:scf}
\end{align}
where $p$ describes the redshift distribution which is essentially the filter profile in units of comoving distance, $R$ is the distance between two sources, and $\bar{r}$ is the mean spatial position of two sources. As discussed in \S\ref{sec:acf}, our samples are not large enough to constrain the slopes of the correlation functions. Therefore, we fix the slopes such that $\beta = -0.8$ and $\gamma = -1.8$ (fiducial\footnotemark; $\beta = \gamma +1$). We use the exact filter profiles associated with each narrow/intermediate band sample and use Equation \ref{eqn:scf} along with our observed measurements of \wtheta~to measure the clustering length, \ro, per each iteration as described in \S\ref{sec:acf}.

\footnotetext{The fiducial slopes were determined by \citet{Totsuji1969} and have remained the same for the past 50 years as it is found to best represent the angular and spatial correlation functions for a wide range of galaxy samples}

The redshift distributions of the combined intermediate samples are the combination of all the filter profiles associated with each respective intermediate band sample. We also weight the redshift distributions by the number of LAEs specific to a intermediate band sample. For example, the $z\sim2.8$ sample consists of 1577 LAEs for which 634, 286, and 657 emitters are from the IA427, IA464, and IA484 samples, respectively, with the 30 percent completeness limits included per sample as reported by \citet{Sobral2018}. Since the number of LAEs is then not homogeneous per intermediate band sample, the final redshift distribution for the combined sample is weighted by the number of emitters in each intermediate sample.

When we measure the clustering properties in bins of galaxy properties (e.g., line luminosity, UV continuum), we are essentially selecting a subsample from the full redshift distribution. Therefore, to properly measure the spatial correlation function, we must augment the weighting of the redshift distributions which is done using the same approach as described above to reflect the relative contribution of each individual intermediate band sample to the combined sample for a given specific bin of galaxy properties. For example, the full $z\sim2.8$ sample has a relative contribution of 40, 18, and 42 percent from the individual IA427, IA464, and IA484-selected LAEs, respectively. If we look at a specific bin in \lya~luminosity that consists of a total of 383 LAEs, the relative contribution changes to 28, 19, and 53 percent for each respective sample. Therefore, we augment the weighted redshift distributions to reflect the changing contributions of each individual intermediate band sample when making our measurements. We follow this approach for all measurements of the spatial correlation function.

\subsection{Dark Matter Halo Model}
\label{sec:dmh}
The spatial clustering of galaxies is related to the overall dark matter distribution as:
\begin{eqnarray}
b^2_{\rm eff} = \frac{\xi_{gg}(r)}{\xi_{mm}(r)}
\end{eqnarray}
where $b_{\rm eff}$ is the effective galaxy bias, $\xi_{gg}$ and $\xi_{mm}$ are the galaxy-galaxy and matter-matter spatial correlation functions, respectively. The effective galaxy bias is related to the halo occupation distribution by:
\begin{eqnarray}
b_{\rm eff}(z) = \frac{\int_{\mathrm{M}_\mathrm{min}}^\infty b_{h}(M,z) n_h(M,z) \langle N_g(M,z) \rangle~ \mathrm{d}M}{\int_{\mathrm{M}_\mathrm{min}}^\infty n_h(M,z) \langle N_g(M,z) \rangle~ \mathrm{d}M}
\label{eqn:dmh}
\end{eqnarray}
where $b_h$ and $n_h$ are the halo bias and mass functions for a given halo mass, $M$, respectively, $N_g(M,z)$ is the galaxy-halo occupation function, and M$_\textrm{min}$ is the minimum dark matter halo mass. The effective halo mass can then be calculated as:
\begin{eqnarray}
M_{\rm eff} = \frac{\int_{\mathrm{M}_\mathrm{min}}^\infty M n_h(M,z) \langle N_g(M,z) \rangle~ \mathrm{d}M}{\int_{\mathrm{M}_\mathrm{min}}^\infty n_h(M,z) \langle N_g(M,z) \rangle~ \mathrm{d}M}
\label{eqn:eff_mass}
\end{eqnarray}
for a given sample at a specific redshift.

There are numerous prescriptions for the galaxy-halo occupation that ranges from simple one-to-one occupation to 3 parameter models (e.g., \citealt{Kravtsov2004}), to 5 parameter models (e.g., \citealt{Zheng2005}), and can be as complex as 12 parameter models (e.g., \citealt{Geach2012}). The one-to-one occupation model assumes a single galaxy resides in each halo, while more complex models take into account multiple galaxies occupying a single halo (central + satellite galaxies). 

Since many of our samples are not large enough to properly constrain multiparameter halo occupation distribution models, we assume a simple one-to-one occupation model ($\langle N_g(M,z) \rangle = 1$) where each LAE is a central galaxy hosted by a dark matter halo by a minimum halo mass, $M_\textrm{min}$. This enables us to adopt a consistent approach throughout this work, even where sample sizes are small. We use the {\sc{Colossus}} package \citep{Diemer2017} in order to measure $\xi_{mm}$ at the redshifts corresponding to our samples. The effective bias is measured at $r = 8\ h^{-1}$ Mpc (comoving) which corresponds to the regime for which the linear matter power spectrum dominates. The \citet{Tinker2010} halo bias and \citet{Tinker2008} halo mass functions are used for $b_h$ and $n_h$ in Equation \ref{eqn:dmh}, respectively.

Throughout this paper, we refer to the effective halo mass as `halo mass' unless otherwise stated.

\subsection{Cosmic (Sample) Variance}
One of the major systematic uncertainties that we take into account in our measurements is the effect of cosmic or sample variance which arises from the limited survey area. \citet{Sobral2010} measured the effects of cosmic variance on the clustering amplitude using their \ha~sample at $z = 0.84$ by randomly sampling areas between 0.05 deg$^2$ to 0.5 deg$^2$ in their 1.3 deg$^2$ coverage of the COSMOS field. They find that the uncertainty in the clustering amplitude as a function of area scales as $20 \times \Omega^{-0.35}$, with $\Omega$ being the survey size in deg$^2$. For the case of our $\sim 2$ deg$^2$ survey (see Table \ref{table:clustering_props} for the survey size per redshift slice), the uncertainty in the clustering amplitude is $\sim 16$ percent of $A_w$, which corresponds to $\sim 11$ percent of the clustering length, \ro. We incorporate these systematic errors by adding them in quadrature to the measured uncertainites.

\subsection{Contamination}

\begin{table*}
	\centering
	\caption{List of the filters corresponding to rest-frame 1500\AA~for each intermediate and narrowband sample. The central wavelength and widths are the rest-frame parameters of the corresponding filters. All photometry used in measuring $\beta_{\rm UV}$ are shown in the $\beta_{\rm UV}$ Filters column with the number of filters used shown as $N_\textrm{filters}$. Note that $\beta_{\rm UV}$ for the combined intermediate band samples is measured as listed in each individual filter. The sample sizes of the combined samples include a 30 percent completeness limit cut on each individual intermediate band sample as measured by \citet{Sobral2018}.}
	\begin{tabular}{cccccclc}
		\hline
		Sample & $z$ & $N_g$ & Filter & Eff. Wave. & FWHM & $\beta_{\rm UV}$ Filters & N$_\textrm{filters}$\\
		&                  & &  &     (\AA)     & (\AA) &  & \\
		\hline
		IA427  & $2.51\pm0.08$ & 748 & $V_{540}$ & 1538.97 & 282.55 & $BVgri$, IA464 -- IA827, NB711, NB816 & 20\\
		IA464  & $2.81\pm0.09$ & 313 & $V_{540}$ & 1414.50 & 260.02 & $VrizY$, IA484 -- IA827, NB711, NB816 & 17\\
		IA484  & $2.99\pm0.09$ & 713 & $r_{645}$ & 1619.17 & 295.38 & $VrizYJ$, IA505 -- IA827, NB711, NB816 & 17\\
		IA505  & $3.17\pm0.09$ & 484 & $r_{645}$ & 1551.84 & 283.10 & $VrizYJ$, IA527 -- IA827, NB711, NB816 & 16\\
		IA527  & $3.33\pm0.10$ & 642 & $r_{645}$ & 1487.06 & 271.28 & $VrizYJ$, IA574 -- IA827, NB711, NB816 & 15\\
		IA574  & $3.74\pm0.11$ &  98 & $i_{790}$ & 1672.97 & 297.20 & $rizYJ$, IA624 -- IA827, NB711, NB816 & 13\\
		IA624  & $4.13\pm0.12$ & 143 & $i_{790}$ & 1538.93 & 273.38 & $izYJ$, IA679 -- IA827, NB711, NB816 & 11\\
		IA679  & $4.58\pm0.14$ &  80 & $z_{915}$ & 1637.86 & 251.20 & $izYJH$, IA709 -- IA827, NB711, NB816 & 11\\
		IA709  & $4.82\pm0.13$ &  63 & $z_{915}$ & 1568.56 & 240.58 & $izYJH$, IA738 -- IA827, NB816 & 9\\
		IA738  & $5.06\pm0.13$ &  79 & $z_{915}$ & 1506.92 & 231.12 & $izYJH$, IA767 -- IA827, NB816 & 8\\
		IA767  & $5.33\pm0.15$ &  33 & $z_{915}$ & 1449.94 & 222.38 & $zYJH$, IA827, NB816 & 6\\
		IA827  & $5.78\pm0.14$ &  36 & $Y_{1029}$ & 1511.68 & 151.97 & $zYJHK$ & 5\\
		\hline
		NB497  & $3.10\pm0.02$ & 1198 & $r_{645}$ & 1576.82 & 287.65 & $grizJ$ & 5\\
		NB711  & $4.86\pm0.03$ &   78 & $z_{915}$ & 1564.14 & 239.90 & $izYJH$, IA738 -- IA827 & 9\\
		NB816  & $5.71\pm0.04$ &  192 & $Y_{1029}$ & 1532.06 & 154.02 & $zYJH$ & 4\\
		\hline
		IA427 - IA484 & $2.75^{+0.33}_{-0.33}$ & 1577 & --- & --- & --- & --- & ---\\
		IA505 - IA527 & $3.25^{+0.18}_{-0.17}$ & 1074 & --- & --- & --- & --- & ---\\
		IA574 - IA624 & $3.94^{+0.32}_{-0.31}$ &  185 & --- & --- & --- & --- & ---\\
		IA679 - IA738 & $4.82^{+0.37}_{-0.38}$ &  192 & --- & --- & --- & --- & ---\\
		IA767 - IA827 & $5.56^{+0.37}_{-0.38}$ &   53 & --- & --- & --- & --- & ---\\
		\hline
	\end{tabular}
	\label{table:muv_filters}
\end{table*}

Contamination is typically assumed to cause an underestimation of the observed clustering signal where contaminants are randomly distributed in the field. Quantitatively, the clustering signal, $A_w$, will be underestimated by a factor of $(1 - f)^2$, with $f$ being the contamination fraction. This translates to a factor of $(1 - f)^{2/|\gamma|}$ for \ro~(the clustering length). 

As discussed in \citet{Khostovan2018}, the effect of contamination is not as straightforward in narrowband samples since the contaminants are other emission line-selected galaxies. In the case of this study, our contaminants will be primarly low-$z$ interlopers, such as \oii, \oiii, and \ha~emitters. These low-$z$ interlopers exhibit non-random clustering (e.g., \citealt{Shioya2008,Sobral2010,Stroe2015,Cochrane2017,Khostovan2018}) and, therefore, can either cause an overestimation or underestimation of the clustering signal. A benefit of SC4K is that \ha~and \oiii~become low-$z$ interlopers starting in IA679 and IA505, respectively, and are small in numbers due to volume such that their effects are negligible. 

\citet{Sobral2018} investigated the contamination fraction for the SC4K sample using the available spectroscopic measurements. Of the 132 sources with spectroscopic redshifts, 112 were confirmed to be LAEs suggesting a contamination fraction of $\sim 15$ percent, which is typical of large-area \lya~narrowband surveys. \citet{Sobral2018} also investigated whether this contamination was dependent on redshift, \lya~luminosity, and rest-frame EW and found that it is constant around 10 - 20 percent. Using the simple $(1-f)^2$ factor, a contamination fraction of 15 percent would increase $A_w$ by $\sim 38$ percent and \ro~by $\sim 20$ percent, but with the assumption that these contaminants are randomly distributed, which, as discussed above, should not be the case. We instead omit from correcting the clustering measurements for contamination effects, but cite the numbers above as the maximum effect contaminants can have on the clustering signal.

\section{Rest-Frame UV Properties}
\label{sec:UV_prop}

\subsection{Determining M$_\textrm{UV}$ and $\beta_{\rm UV}$}
\label{sec:MUV}

The typical shape of the SED of star-forming galaxies at $1300\textrm{\AA} < \lambda < 2800\textrm{\AA}$ can be fit by a power law of the form $f_\lambda \propto \lambda^{\beta_{\rm UV}}$, where $f_\lambda$ is the flux density, typically in units of erg s$^{-1}$ cm$^{-2}$ \AA$^{-1}$, and $\beta_{\rm UV}$ is the UV spectral slope. Since the cross-section of dust grains effectively absorbs UV light, the amount of dust attenuation can be measured using the UV slopes (e.g, \citealt{Calzetti1994,Meurer1999}), although it should be noted that redder UV slopes ($\beta_{\rm UV} > -2$) can also signify galaxies with mature, evolved stellar populations. We expect this degeneracy in interpreting $\beta_{\rm UV}$ to be negligible as our samples are emission line-selected and are then typically dominated by populations of star-forming galaxies.

We measure $\beta_{\rm UV}$ by fitting the power law described above using the available photometry in the rest-frame range of $1300\textrm{\AA} < \lambda < 2800\textrm{\AA}$. We measure the 1500 \AA~UV continuum absolute magnitudes (\muv) by:
\begin{equation}
\centering
{\textrm \muv} = m_{\rm UV} - 5\log_{10}\Bigg(\frac{d_L}{10{\rm pc}}\Bigg) + 2.5\log_{10}(1+z)
\end{equation}
where $m_{\rm UV}$ is the observed UV magnitude and $d_L$ is the luminosity distance. Table \ref{table:muv_filters} shows the corresponding observer-frame photometric band associated with $m_{\rm UV}$ used to measure \muv, the rest-frame effective wavelength and FWHM of the filter, and the observer-frame filters used in measuring $\beta_{\rm UV}$. Although the filters cover 1500\AA~within the FWHM, their effective wavelengths are off-centered by a maximum of $\sim 170$\AA~which causes an offset in our measurements of \muv. We calculate the maximum offsets to be $\sim -0.06$, $-0.12$, and $-0.18$ mag for UV spectral slopes of $\beta = -1.5$, $-1$, $-0.5$ and for a $170$\AA~offset towards redder wavelengths. In principle, the offset can be taken into account by applying the correction, $-2.5 (\beta_{\rm UV}+2)\log_{10}(\lambda/1500\textrm{\AA})$, to \muv. Since our LAEs typically have blue spectral slopes, especially for the higher redshift samples, the offsets are negligible. For the case of our sources with redder spectral slopes ($\beta_{\rm UV} > -2$), we find that the uncertainties in \muv~are larger than the offsets.

Since our samples are \lya-selected, we are prone to detect low stellar mass sources for which the stellar continuum is below the survey detection limit compared to continuum-selected surveys. In such cases, we apply a lower limit to \muv~by using the $3\sigma$ detection limit of the photometry. The lack of stellar continuum also means that we are not able to measure $\beta_{\rm UV}$ for a subset of our sources. There are also sources for which the uncertainties in $\beta_{\rm UV}$ are quite high due to weak stellar continuum measurements. To take these effects into account, we take all measurements of $\beta_{\rm UV}$ that have a S/N ($\mathopen|\beta_{\rm UV}/\Delta\beta_{\rm UV}\mathclose|$) $ > 3$ and measure the median. This is then used as the median stacked spectral slope for those that have S/N $< 3$ or no $\beta_{\rm UV}$ measurement.

\subsection{Star Formation Rates of LAEs}

Typically, narrowband surveys measure star formation rates using the observed or dust-corrected emission line luminosity in conjunction with a star formation rate calibration (e.g., \citealt{Ly2011,Sobral2013,Sobral2014}). In the case of LAEs, measuring star formation rates using the \lya~line introduces several caveats. Even though \lya~traces the ionizing radiation of star formation activity, it is easily scattered. This increases the likelihood of being absorbed by dust and also decreases the surface brightness, such that \lya~photons are spread out over a larger area (see \citealt{Dijkstra2017} and \citealt{SM2018} for discussion).

To measure the star formation rates of our LAEs, we instead use the UV continuum luminosities, \muv, as described in \S\ref{sec:MUV}, which traces the population of short-lived, massive \textit{O}, \textit{B}, and \textit{A} type stars corresponding to a star formation activity timescale of $\sim 100$ Myr. We assume the \citet{Kennicutt1998} SFR(UV) calibration:
\begin{equation}
\label{eqn:sfr}
\textrm{SFR(UV)} = 1.4\times10^{-28} \Bigg(\frac{L_\nu}{\mathrm{erg}~ \mathrm{s}^{-1} ~\mathrm{Hz}^{-1}}\Bigg)~\mathrm{M}_\odot~\mathrm{yr}^{-1}
\end{equation}
where $L_\nu$ is the UV luminosity per unit frequency. This calibration is valid for the range of 1500 \AA~to 2800 \AA, where $L_\nu$ is consistently flat (assuming $\beta_{\rm UV} \sim -2$) and assumes a Salpeter IMF. We assume the \citet{Meurer1999} calibration to dust correct \muv:
\begin{equation}
A_\textrm{UV} = 4.43 + 1.99\beta_{\rm UV}
\end{equation}
where $A_\textrm{UV}$ is the UV dust extinction and $\beta_{\rm UV}$~is the UV spectral slope described in \S\ref{sec:MUV}.

\section{Results \& Discussion}
\label{sec:results}

\subsection{Clustering Properties of LAEs}

\begin{figure}
	\centering
	\includegraphics[width=\columnwidth]{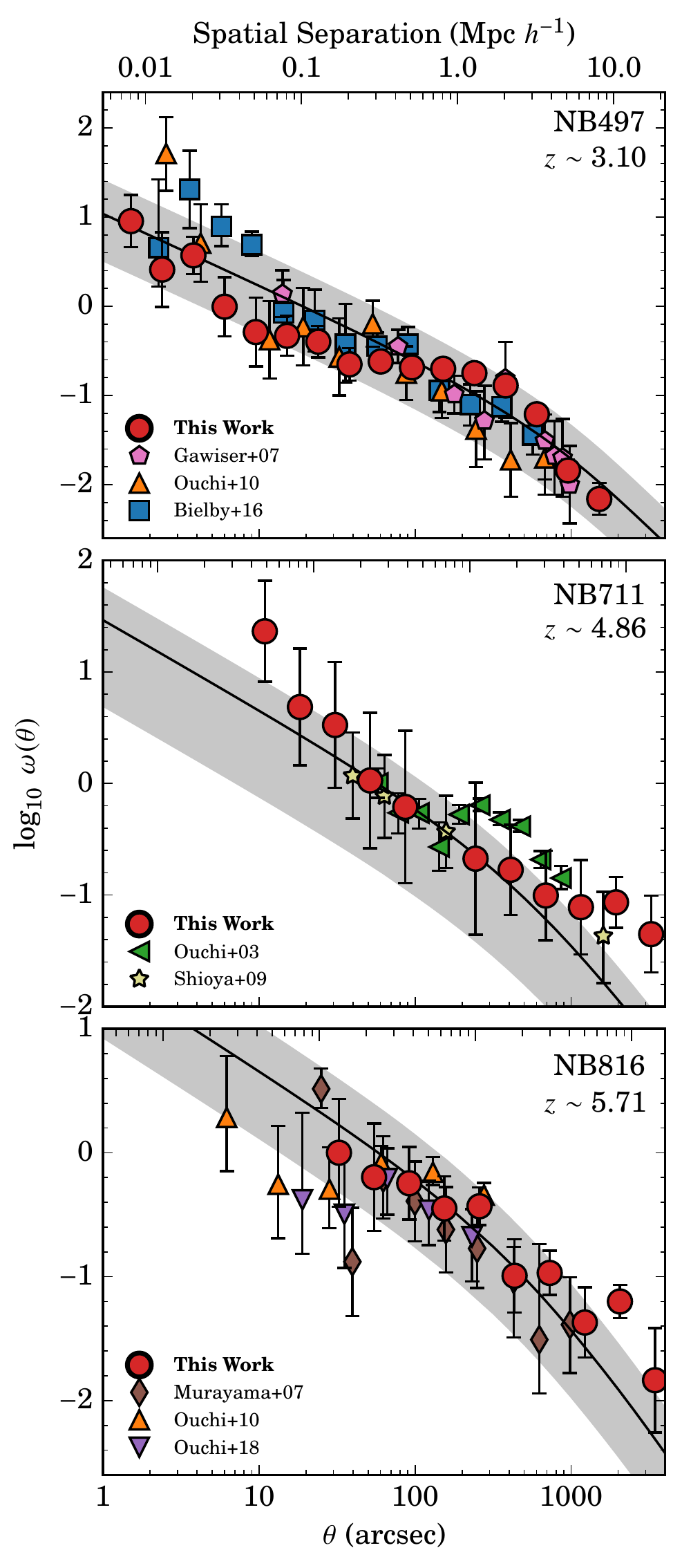}
	\caption{The angular correlation functions for each narrowband sample. The {\it red circles} are the median observed measurements for \wtheta~based on all 2000 iterations of measuring \wtheta~with varying bin sizes and centers. The {\it black line} shows the best-fit model as described in Equation \ref{eqn:scf} with the $1\sigma$ uncertainty represented as the {\it grey}. The spatial axis shown in each panel corresponds to the spatial separation for a given angular separation at the redshift of the samples shown. We only detect the $1-$halo term at $\theta < 10''$ ($r < 53$ kpc $h^{-1}$) for our $z\sim3.1$ sample, which suggests that our sample does contain satellites. We ruled out the large overdensity in the SA22 field as the source of the $1-$halo term (see Appendix \ref{sec:overdense}). Overlaid are the ACFs from various narrowband surveys that are consistent with the redshifts of our samples.} 
	\label{fig:acf_NB}
\end{figure}

In this section, we present our results on the clustering properties of our LAE samples. The angular correlation functions are shown for all our narrowband, intermediate band, and combined samples in Figures \ref{fig:acf_NB}, \ref{fig:acf_IB}, and \ref{fig:acf_combined}, respectively. Each bin in angular separation encompasses the median measurement of \wtheta~for all 2000 realizations and the errors includes the scatter in \wtheta~and the median Poisson error described in Equation \ref{eqn:errors}. In this respect, we are taking into account the effects of selecting some arbitrary fixed bin size and bin width in making our final ACF measurements. We overlay the fits based on the measured \ro~in Equation \ref{eqn:scf} and shown in Table \ref{table:clustering_props} and find that it is consistent with the median ACFs. Note that, as described above, the measured \ro~is based on the distribution of 2000 \ro~measurements that correspond to each individual ACF.

As described in \S\ref{sec:acf}, we use the exact form of the Limber Equation as outlined by \citet{Simon2007} to fit the ACFs and find that it best represents the observed measurements, especially at higher angular separations where the deviation from a simple power law occurs. This is more pronounced in the $z \sim 3.1$ NB497 ACF at angular separations greater than $600''$, which corresponds to comoving separations of $\sim 3.4$ Mpc $h^{-1}$, as shown in the top panel of Figure \ref{fig:acf_NB}. Previous narrowband studies have also observed deviations from the simple power law form at high angular separations (e.g., \citealt{Sobral2010,Cochrane2017,Khostovan2018}). 

Figure \ref{fig:acf_NB} includes the observed ACFs from various clustering studies of LAEs \citep{Ouchi2003,Gawiser2007,Murayama2007,Shioya2009,Ouchi2010,Bielby2016,Ouchi2018}. We find that, in general, our ACFs and fits are in agreement with the literature. Note that the ACFs shown are not corrected for cosmic variance effects, which would severely affect $\sim$ arcmin$^2$ measurements such as those of \citet{Ouchi2003} and \citet{Gawiser2007} (see Table \ref{table:lit_r0} for survey size). This is probably why the $z \sim 4.86$, 543 arcmin$^2$ Subaru Deep Survey measurements of \citet{Ouchi2003} are systematically above our measurements, but still within 1$\sigma$.

\subsubsection{Effects of the 1-halo Term}
Typically, ACFs trace two distinct clustering regimes. The first is the galaxy-galaxy angular correlation within a single dark matter halo, referred to as the 1-halo term. The second is the galaxy-galaxy angular correlation, with galaxies residing in separate dark matter halos, which is referred to as the 2-halo term. The 1-halo term signal is observed at low angular separations as a deviation from a simple power law ACF model towards higher \wtheta~(enhanced clustering) and traces the clustering properties of both central and satellite galaxies, while the 2-halo term is observed at larger angular separations. 

We find that most of our samples show no significant detection of the 1-halo term, which suggests that the LAEs in our sample are primarily centrals and have low/negligible satellite fractions. This could be due to selection bias as we are selecting LAEs with strong emission lines and are missing the faint, low-mass population that forms the bulk of the satellite population.

We detect a signature of a $1-$halo term in the $z\sim 3.1$ NB497 sample at angular (comoving) separations of $\sim 10''$ ($\sim 50$ kpc $h^{-1}$), although we note that the observed ACFs are still consistent with the exact Limber equation fits. One possible reason for the detection of the $1-$halo term could be the presence of a significant overdense region in the SA22 field \citep{Steidel1998,Steidel2000,Matsuda2004,Yamada2012} which, in principle, would cause elevated correlation function measurements at lower angular separations. We test this idea in Appendix \ref{sec:overdense} by masking the overdense region and repeating our measurements. We find no significant difference between the ACFs for the full SA22 field and the case where the overdense region is masked.

Another possibility is that the $z \sim 3.1$ NB497 sample is deep enough in line luminosity to observe satellite LAEs. We test this idea in Appendix \ref{sec:overdense} by applying varying line luminosity thresholds and find that the $1-$halo term disappears at $L_\textrm{\lya} \gtrsim 0.4\ L^\star$, such that the satellite fraction is negligible beyond this threshold. Other clustering studies of emission line galaxies \citep{Cochrane2017} and LBGs \citep{Harikane2018} also show that the satellite fractions are typically $\lesssim 5$ percent, such that they are negligible. Since measurements of the satellite population is not the main focus of this paper, we defer further discussion but assume based on past works and our own observations that such a population has minimal effects on our measurements, particularly for the $L_\textrm{\lya} \gtrsim 0.4\ L^\star$ population.

\begin{figure*}
	\centering
	\includegraphics[width=\textwidth]{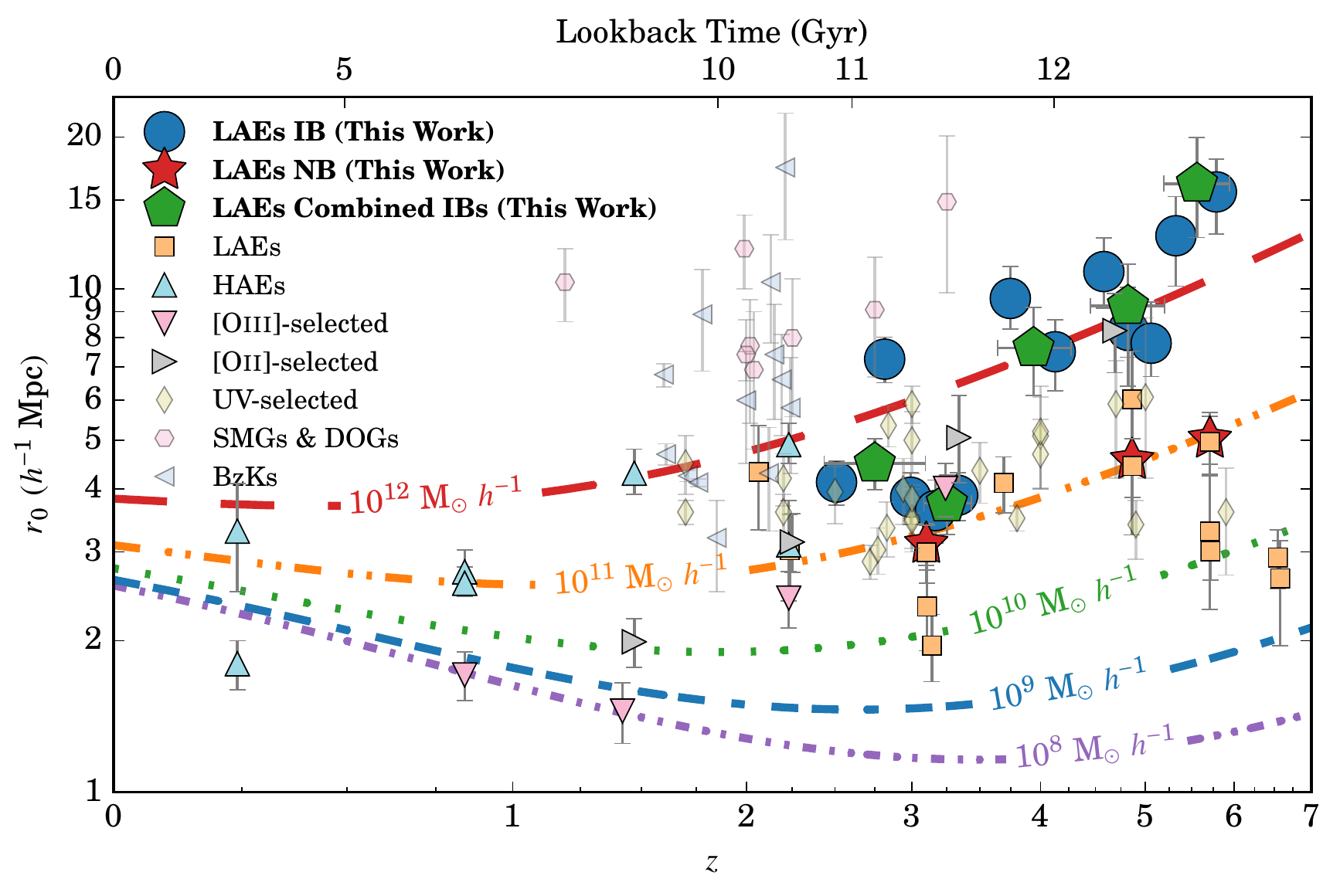}
	\caption{The redshift evolution of the clustering length, \ro, for our intermediate band, combined, and narrowband-selected \lya~emitters. We find that \ro~increases with increasing redshift up to $z\sim6$ for all our samples. The systematic offset in \ro~between our intermediate and narrowband-selected samples is attributed to selection effects (e.g., depth of each sample). Comparison with narrowband-selected samples drawn from the literature also show widely varying \ro \citep{Ouchi2003,Gawiser2007,Murayama2007,Shioya2009,Guaita2010,Ouchi2010,Bielby2016,Ouchi2018}. The redshift evolutions of host dark matter halos are also shown for minimum halo masses between $10^{8 - 12}$ \msol~$h^{-1}$. Our intermediate band-selected LAEs are found to be hosted by $10^{11 - 12}$ \msol~halos, while our narrowband-selected LAEs are hosted by $\sim 10^{11}$ \msol~halos for all redshifts. We also compare to \ha~\citep{Geach2008,Shioya2008,Sobral2010,Stroe2015,Cochrane2017}, \oii~\citep{Khostovan2018}, and \oiii~\citep{Khostovan2018} narrowband studies and find that, for the overlapping redshift ranges, we are in agreement, suggesting that the various emission line-selected galaxies reside in dark matter halos with similar masses. Included are measurements from UV-selected samples \citep{Foucaud2003,Ouchi2004,Adelberger2005,Kashikawa2006,Savoy2011,Bielby2013,Barone2014,Durkalec2015,Durkalec2018}, Sub-millimeter Galaxies (SMGs; \citealt{Blain2004,Hickox2012,Wilkinson2017}), Dust-Obscured Galaxies (DOGs; \citealt{Brodwin2008,Toba2017}), and star-forming $BzK$-selected galaxies ($BzK$s; \citealt{Kong2006,Hayashi2007,Blanc2008,Hartley2008,McCracken2010,Lin2012,Ishikawa2015}). Each selection type covers a specific subset of star-forming galaxies resulting in wide ranges of clustering lengths, which highlights the need to properly compare samples by accounting for sample biases.}
	\label{fig:r0}
\end{figure*}

\subsubsection{Clustering Length}
With the observed ACFs, we measure the spatial clustering lengths using our approach highlighted in \S\ref{sec:spatial_corr}. Figure \ref{fig:r0} shows the redshift evolution of the clustering length, \ro, for all our LAE samples without any cuts. Although there is a distinct difference between the NB- and IB (combined IB)-selected results, both show an increasing \ro~with increasing redshift and are shown in Table \ref{table:clustering_props}. Comparing the NB- and IB-samples yields an apparently different redshift evolution where by $z = 5.7$, the IB-samples are three times as clustered compared to the NB-samples. A similar result is seen when comparing the NB- and combined IB-samples. The main cause of this difference is due to sample selection and survey parameters as the narrowbands are $\sim 0.2 - 0.8$ dex deeper in \lya~luminosity than their corresponding intermediate bands, whilst covering a smaller volume (see Table \ref{table:clustering_props}).

The issue of sample selection effects on the clustering results become evident when comparing IB-to-IB samples. For example, the $z = 2.8$ IA464 sample is $0.2$ and $0.4$ dex shallower in depth in comparison to the $z = 2.5$ IA427 and $z = 3.0$ IA484 sample, respectively, and is found to be more clustered by a factor of two. This suggests that the clustering signal is dependent on \lya~luminosity and to properly compare clustering properties requires that we take this effect into account. We showed the importance of this effect in an earlier work for other emission line-selected samples \citep{Khostovan2018}.

Figure \ref{fig:r0} also includes the clustering lengths associated with minimum halo masses between $10^{8 - 12}$ \msol~host dark matter halos as a function of redshift. We find that our IB-selected LAEs typically reside in host halos with a minimum mass range of $\sim 10^{11 - 12}$ \msol~and the NB-selected LAEs show a consistent minimum host halo mass of $\sim 10^{11}$ \msol~for all redshifts observed. There is no redshift evolution observed in host halo mass, which suggests that galaxies observed as LAEs at different redshifts reside in halos of similar mass. Table \ref{table:clustering_props} highlights the effective host halo masses per sample.

\begin{table*}
	\centering
	\caption{The clustering properties for the full population of LAEs per sample. Shown are the redshifts, filter names, number of LAEs per sample, the corresponding survey area and comoving volume in deg$^2$ and Mpc$^3$, respectively, the characteristic line luminosity ($L^\star(z)$), the median \lya~luminosity, the clustering amplitude measured from the observed ACFs, the exact clustering length, \ro, measured using Equation \ref{eqn:scf}, and the effective halo mass measured using our model described in \S\ref{sec:dmh}. Each sample presented is within the $\sim 2$ deg$^2$ COSMOS field, except for the $1.38$ deg$^2$ SA22 NB497 sample. All $L^\star(z)$ measurements are from \citet{Sobral2018} except for the narrowband samples. The NB816 $L^\star$ is measured by \citet{Santos2016}. We use the redshift evolution of $L^\star(z)$ from the SC4K samples measured in \citet{Sobral2018} to measure $L^\star(z)$ for the NB497 and NB711 samples. Our NB711 $L^\star(z)$ measurement is consistent with \citet{Shioya2009} which measured $\log_{10} L^\star = 42.9^{+0.5}_{-0.3}$ erg s$^{-1}$.}
	\label{table:clustering_props}
	\begin{tabular*}{\textwidth}{@{\extracolsep{\fill}}ccccccccccc}
		\hline
		$z$ & Filter & $N_\textrm{gal}$ & Area & Volume & $\log_{10}\ L^\star(z)$ & Med. $\log_{10}\ L$ &$A_w$ & $r_0^\textrm{exact}$ & $\log_{10}$ Halo Mass \\
		&  &  & (deg$^2$) & ($10^6$ Mpc$^3$) & (erg s$^{-1}$) & (erg s$^{-1}$) & (arcsec$^{-0.8}$) & (Mpc $h^{-1}$)  & (M$_\odot$ $h^{-1}$) \\
		\hline
		2.51     & IA427 & 748  & 1.94 & 4.0 & 42.76$^{+0.07}_{-0.07}$ & 42.70 & $6.47^{+1.25}_{-1.20}$     & $4.13^{+0.42}_{-0.42}$  & $11.59^{+0.16}_{-0.16}$ \\
		2.81     & IA464 & 313  & 1.94 & 4.2 & 42.83$^{+0.36}_{-0.19}$ & 43.06 & $15.51^{+3.10}_{-2.93}$   & $7.24^{+0.76}_{-0.74}$  & $12.28^{+0.13}_{-0.13}$ \\
		2.99     & IA484 & 713  & 1.94 & 4.3 & 42.64$^{+0.06}_{-0.05}$ & 42.86 & $4.86^{+1.13}_{-1.08}$    & $3.85^{+0.46}_{-0.45}$  & $11.30^{+0.20}_{-0.19}$ \\
		3.17     & IA505 & 484  & 1.94 & 4.3 & 42.80$^{+0.09}_{-0.07}$ & 42.92 & $4.46^{+1.42}_{-1.09}$    & $3.62^{+0.54}_{-0.50}$  & $11.14^{+0.25}_{-0.23}$ \\
		3.33     & IA527 & 642  & 1.94 & 4.5 & 42.68$^{+0.07}_{-0.06}$ & 42.86 & $5.94^{+1.13}_{-1.13}$    & $3.89^{+0.43}_{-0.42}$  & $11.20^{+0.18}_{-0.18}$ \\
		3.74     & IA574 &  98  & 1.96 & 4.9 & 43.03$^{+0.18}_{-0.15}$ & 43.12 &  $23.39^{+7.97}_{-6.13}$   & $9.56^{+1.50}_{-1.26}$  & $12.31^{+0.19}_{-0.16}$ \\
		4.13     & IA624 & 143  & 1.96 & 5.2 & 42.83$^{+0.17}_{-0.15}$ & 43.02 & $15.59^{+5.09}_{-4.68}$   & $7.49^{+1.17}_{-1.22}$  & $11.89^{+0.20}_{-0.21}$ \\
		4.58     & IA679 &  80  & 1.96 & 5.5 & 43.15$^{+0.16}_{-0.15}$ & 43.26 & $37.35^{+10.52}_{-10.51}$ & $10.81^{+1.79}_{-1.66}$ & $12.21^{+0.20}_{-0.18}$ \\
		4.82     & IA709 &  63  & 1.96 & 5.1 & 42.98$^{+0.17}_{-0.14}$ & 43.22 & $24.38^{+11.39}_{-8.69}$  & $8.26^{+1.79}_{-1.87}$  & $11.81^{+0.28}_{-0.29}$ \\
		5.06     & IA738 &  79  & 1.96 & 5.1 & 43.30$^{+0.23}_{-0.19}$ & 43.42 &  $19.68^{+8.13}_{-5.54}$   & $7.79^{+1.61}_{-1.10}$  & $11.67^{+0.27}_{-0.19}$ \\
		5.33     & IA767 &  33  & 1.96 & 5.5 & 43.30$^{+0.28}_{-0.20}$ & 43.55 &  $39.53^{+19.24}_{-18.98}$ & $12.74^{+2.50}_{-2.62}$ & $12.21^{+0.23}_{-0.24}$ \\
		5.79     & IA827 &  36  & 1.96 & 4.9 & 43.35$^{+0.24}_{-0.19}$ & 43.60 & $76.99^{+25.06}_{-24.01}$ & $15.56^{+2.51}_{-2.71}$ & $12.34^{+0.18}_{-0.20}$ \\
		\hline
		3.10     & NB497 & 1198 & 1.38 & 1.0 & 42.77 & 42.22 & 
		$8.95^{+1.54}_{-1.55}$ & $3.11^{+0.30}_{-0.29}$ &
		$10.89^{+0.18}_{-0.17}$ \\
		4.86     & NB711 & 78   & 1.96 & 1.2 & 43.15 & 43.05 &
		$17.85^{+10.81}_{-7.43}$  & $4.57^{+1.24}_{-1.33}$ & $10.97^{+0.42}_{-0.45}$ \\
		5.71     & NB816 & 172  & 1.96 & 1.8 & 43.25$^{+0.09}_{-0.06}$ & 42.82 & $19.18^{+4.07}_{-3.83}$   & $5.04^{+0.55}_{-0.56}$  & $10.87^{+0.17}_{-0.17}$ \\
		\hline
		2.75     & --- & 1577 & 1.94 & 12.5 & --- & --- & $2.89^{+0.63}_{-0.59}$    & $4.50^{+0.54}_{-0.51}$  & $11.63^{+0.18}_{-0.17}$ \\
		3.25     & --- & 1074 & 1.94 & 8.8 & --- & --- & $3.17^{+0.85}_{-0.74}$    & $3.75^{+0.52}_{-0.50}$  & $11.17^{+0.23}_{-0.22}$ \\
		3.94     & ---      & 185  & 1.96 & 10.1 & --- & --- & $10.41^{+4.26}_{-3.52}$   & $7.62^{+1.56}_{-1.48}$  & $11.97^{+0.26}_{-0.25}$ \\
		4.82     & ---  & 192  & 1.96 & 15.7 & --- & --- & $12.71^{+4.36}_{-3.98}$   & $9.24^{+1.94}_{-1.93}$  & $11.96^{+0.26}_{-0.26}$ \\
		5.56     & ---       & 53   & 1.96 & 10.4 & --- & --- & $44.55^{+19.63}_{-15.12}$ & $16.16^{+3.80}_{-3.52}$ & $12.43^{+0.26}_{-0.24}$\\
		\hline
	\end{tabular*}
\end{table*}

We also include the \ro~measurements of NB-selected LAEs drawn from the literature in Figure \ref{fig:r0} \citep{Ouchi2003,Murayama2007,Guaita2010,Ouchi2010,Bielby2016,Ouchi2018}. Differences in measuring clustering lengths and halo masses in comparison to our approach are taken into account and described in Appendix \ref{sec:literature}. Figure \ref{fig:r0} also includes \ha~\citep{Shioya2008,Sobral2010,Stroe2015,Cochrane2017,Kashino2017}, \oiii~\citep{Khostovan2018}, and \oii~emitters \citep{Takahashi2007, Khostovan2018}.

We find an excellent agreement between our measurement of $\textrm{\ro} = 3.11^{+0.30}_{-0.29}$ Mpc $h^{-1}$ for our $z\sim3.1$ NB497 sample and \ro~$=2.99\pm0.40$ Mpc $h^{-1}$ from the VLT LBG redshift survey of \citet{Bielby2016}. Both our work and \citet{Bielby2016} use a similar NB497 filter and are somewhat similar in survey parameters and selection, although their sample size is smaller ($\sim 600$ LAEs) and they apply a higher rest-frame equivalent width cut (65\AA). We find that the other $z \sim 3.1$ studies report a lower \ro~with a $>1\sigma$ deviation with the ECDF-S MUSYC imaging survey of \citet{Gawiser2007} measuring an $\textrm{\ro} =  2.34\pm 0.43$ Mpc $h^{-1}$ and \citet{Ouchi2010} measuring $\textrm{\ro} =  1.96\pm0.30$ Mpc $h^{-1}$ for LAEs in the SXDS field. The \citet{Ouchi2010} $z \sim 3.1$ sample is somewhat deeper than our NB497 sample with a limiting flux of $\sim1.2\times10^{-17}$ erg s$^{-1}$ cm$^{-2}$. The \citet{Gawiser2007} sample is also somewhat deeper with a limiting flux of $\sim 1.5 \times 10^{-17}$ erg s$^{-1}$ cm$^{-2}$, such that the discrepancy is most likely due to the fainter LAEs being picked up by the two respective studies.

Our $z = 4.86$ NB711 \ro~measurement is found to be in agreement with the \citet{Shioya2009} $1.83$ deg$^{2}$ COSMOS measurement of $\textrm{\ro} = 4.44\pm0.59$ Mpc $h^{-1}$, despite the different source extraction and sample selection used by \citet{Sobral2018}. \citet{Ouchi2003} performed an earlier clustering analysis of LAEs in the $543$ arcmin$^{2}$ Subaru Deep Field using a similar NB711 filter and reported a $\textrm{\ro} = 6.03\pm1.49$ Mpc $h^{-1}$, which is within $1\sigma$ agreement with our results. 

We also find an agreement within $1\sigma$ between our $z = 5.71$ NB816 \ro~measurement and that of the $1.95$ deg$^2$ COSMOS measurement of \citet{Murayama2007}, while the SXDS measurements of \citet{Ouchi2010} and the HSC SILVERRUSH measurements of \citet{Ouchi2018} are lower. The cause of the difference is likely due to survey depth (\lya~luminosity; e.g., the SXDS measurements are close to 1 mag deeper in terms of 5$\sigma$ narrowband detection limits) and also cosmic variance. 

Comparing our measurements to continuum-selected samples from the literature shows that \lya~emitters and LBGs have similar \ro, as shown in Figure \ref{fig:r0}.  Measurements from $BzK$, SMG, and DOG-selected samples show clustering lengths higher than our \lya~measurements at all redshifts. This is primarily due to selection effects as samples, such as DOGs, will select more massive, dustier populations in comparison to our LAE samples, which primarily select dust-free, low mass systems.

\subsection{Line Luminosity}
\label{sec:lum}

\begin{figure}
	\centering
	\includegraphics[width=\columnwidth]{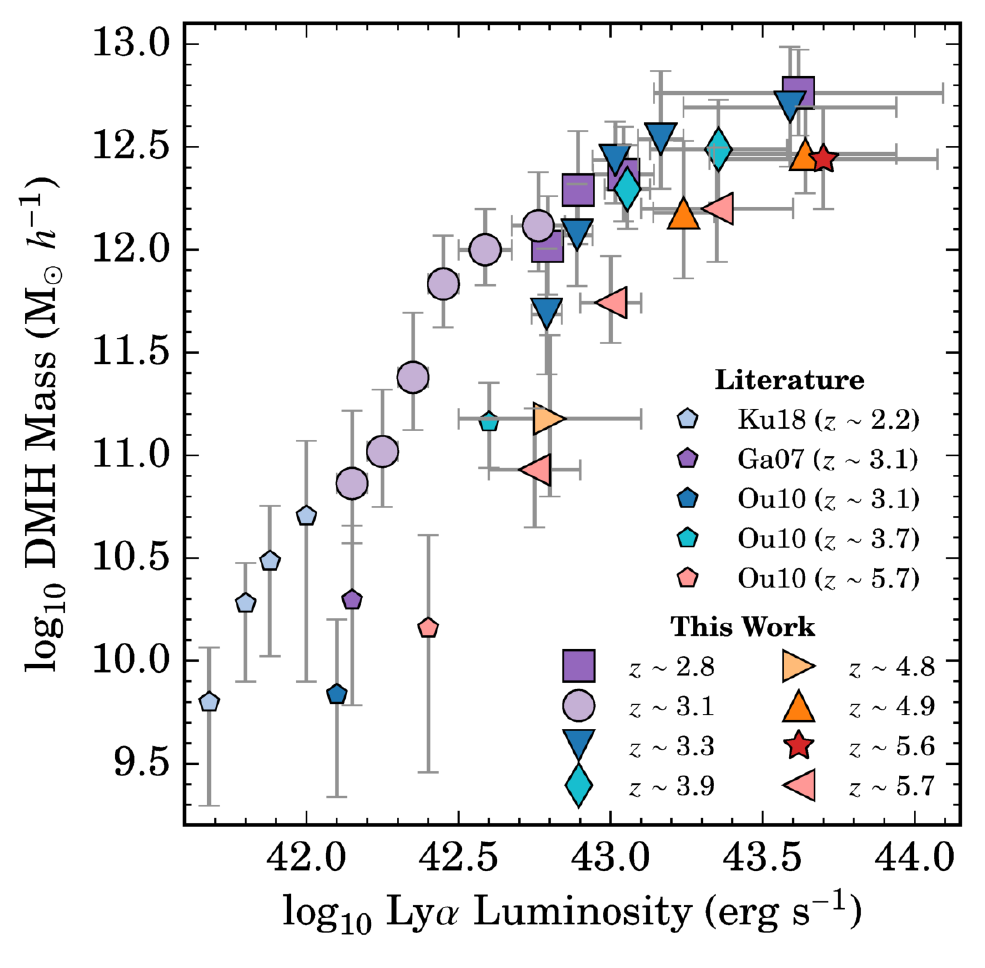}
	\caption{Halo mass in terms of observed \lya~luminosity. For each redshift sample, we see that halo mass increases with increasing line luminosity. Between $z \sim 2 - 3$, our measurements show an increase in halo mass from $10^{9.7 - 12.8}$ \msol~for \lya~luminosities between $10^{41.7 - 43.6}$ erg s$^{-1}$. Similar trends are also seen at $z > 3$, but are shifted to higher line luminosities in comparison to the $z \sim 2 - 3$ samples.}
	\label{fig:halo_lum_only}
\end{figure}

\begin{figure}
	\centering
	\includegraphics[width=\columnwidth]{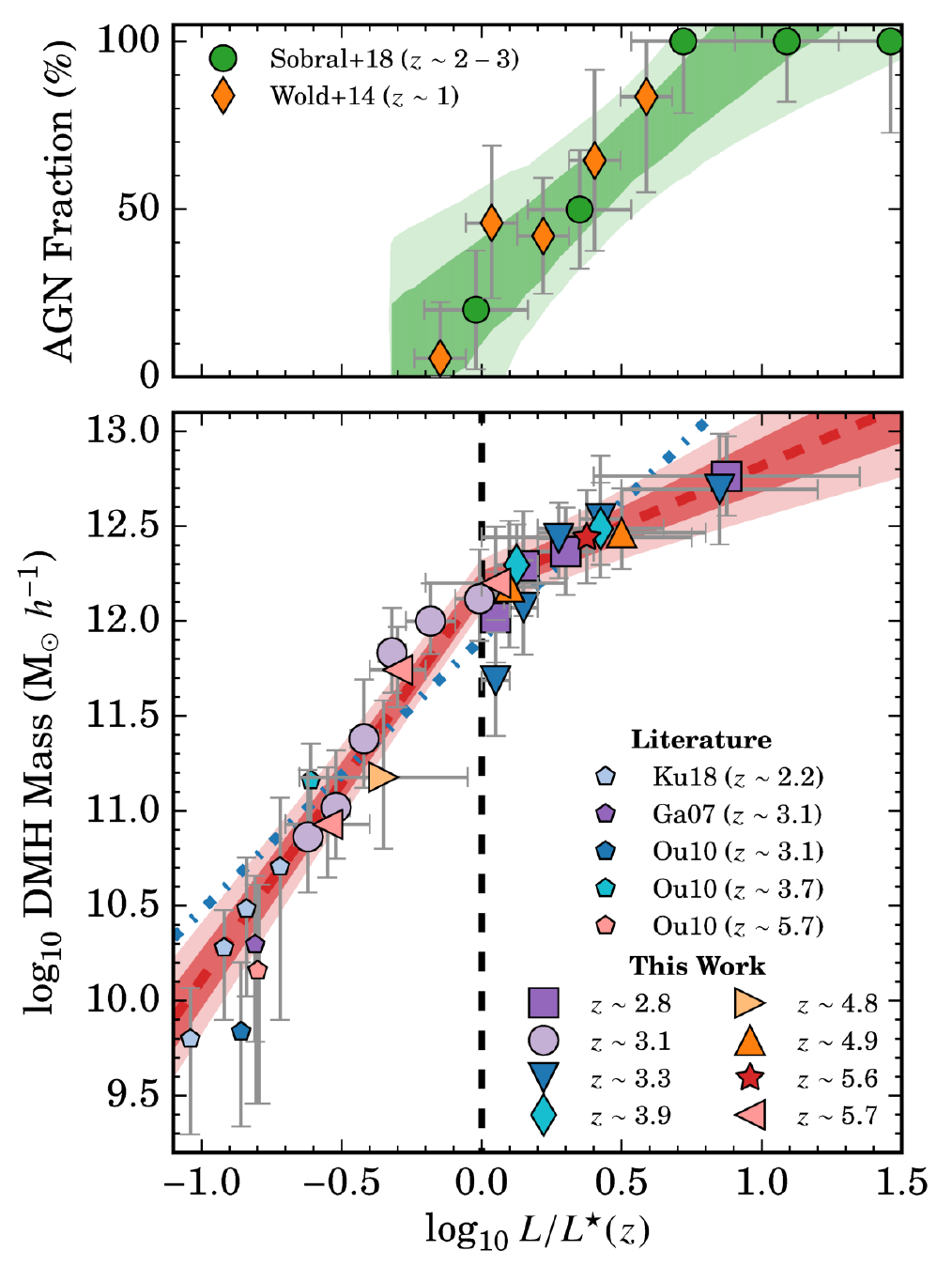}
	\caption{Host halo mass and \lya~luminosity normalized by the characteristic line luminosity, $L^\star(z)$. We find a strong, redshift-independent trend between host halo mass and the $L^\star(z)$ normalized line luminosity similar to previous narrowband works for \ha-, \oii-, and \oiii-selected emission line galaxies \citep{Sobral2010,Cochrane2017,Khostovan2018}. These are quantified by a single ({\it blue line}) and piecewise ({\it red line}) power law model, with the $1\sigma$ and $2\sigma$ regions shown in {\it dark} and {\it light red}, respectively. We find that the observed trends become shallower at $L>L^\star(z)$, which may be a signature of a transitional halo mass where it becomes increasingly improbable that a star-forming galaxy resides in higher host halo masses. The continuous, shallower increase can also be a sign of AGN contribution at the brightest \lya~luminosities. Recent work by \citet{Sobral2018_AGN} find a strong increase in the AGN fractions with $L/L^\star(z)$ and that by $L \sim 2L^\star(z)$, the AGN fraction is $\approx$ 50 percent.}
	\label{fig:lum}
\end{figure} 

Motivated by the results of \citet{Sobral2010}, \citet{Cochrane2017}, and \citet{Khostovan2018}, we investigate the trends between \lya~luminosity and host dark matter halo mass. Our measurements reported here are the first in the literature that cover a wide dynamic range in \lya~luminosity and redshift. Throughout the rest of this paper, we will use our combined intermediate band samples, along with our narrowband samples, to increase the sample statistics, while also ensuring that the redshift range per sample is small enough that any redshift evolution within each combined sample is negligible.

We show the host halo mass in terms of \lya~luminosity in Figure \ref{fig:halo_lum_only}. For all redshift samples, we find that host halo mass increases with increasing \lya~luminosity. Including the \lya~luminosity threshold $z = 2.2$ \citep{Kusakabe2018} and $z = 3.1$ \citep{Gawiser2007,Ouchi2010} literature measurements, along with our $z \sim 2 - 3$ samples, we find that halo mass increases from $10^{9.7 - 12.8}$ \msol~between $10^{41.7 - 43.6}$ erg s$^{-1}$ in \lya~luminosity. We find similar results when looking at the higher redshift samples, in conjunction with luminosity threshold measurements from the literature. The main difference between the redshift samples is an offset in \lya~luminosity with the high-$z$ measurements shifted to higher \lya~luminosities. This could be due to the cosmic evolution in the \lya~luminosity functions. If so, this could be taken into account in order to investigate the evolution of clustering/halo properties of LAEs.

Figure \ref{fig:lum} shows host halo mass in terms of \lya~luminosity normalized by the characteristic line luminosity, $L^\star(z)$. The measurements of $L^\star(z)$ used are shown in Table \ref{table:clustering_props} and are taken from \citet{Sobral2018}, which used the same SC4K sample we use in this study. Since we combine our intermediate band samples, we carefully take into account the variation in $L^\star(z)$ between each individual intermediate band sample by first applying the corresponding $L^\star(z)$ and then binning in terms of $L/L^\star(z)$.

We find a strong, increasing trend between host halo mass and $L/L^\star(z)$ from $z \sim 2.8 - 5.7$, covering 1.3 Gyr of cosmic history from the end of reionization to the peak of cosmic star formation. Our faintest LAEs ($L \sim 0.25 L^\star(z)$) are observed to reside in $10^{10.8}$ \msol~halos and our brightest LAEs ($L \sim 7 L^\star(z)$) reside in $10^{12.8}$ \msol~halos. The typical $L^\star$ galaxy is observed to be found in $\sim 10^{12}$ \msol~host dark matter halos. Surprisingly, these are found to be redshift-independent suggesting that LAEs of the same $L^\star(z)$ type at different redshifts reside in similar halo masses.

Figure \ref{fig:lum} also includes \lya~luminosity threshold measurements drawn from the literature at various redshifts \citep{Gawiser2007,Ouchi2010,Kusakabe2018}. Due to the nature of these measurements, they help to constrain the faint-end of Figure \ref{fig:lum} and are primarily single measurements per redshift, except for \citet{Kusakabe2018}, which made five measurements (although we only show four as their deepest measurement is poorly constrained). The literature measurements, along with our own observations, show significantly strong, redshift-independent trends between \lya~luminosity and effective halo mass. 

To quantify the trends seen in Figure \ref{fig:lum}, we fit two different models: a single power law and a piecewise power law with the pivot point at $L^\star$. The best-fit single power law is:
\begin{eqnarray}
\frac{\textrm{M}_\textrm{halo}}{\textrm{M}_\odot/h} =  10^{11.91^{+0.05}_{-0.05}} \Bigg(\frac{L}{L^\star(z)}\Bigg)^{1.44^{+0.14}_{-0.12}}
\end{eqnarray}
with a slope near unity. Although the single power law seems to represent the observations around $L \sim L^\star(z)$, there is a deviation towards lower and higher line luminosities. Based on this deviation, we use a piecewise power law that is separated at $L^\star(z)$ with a best-fit of:
\begin{eqnarray}
\frac{\textrm{M}_\textrm{halo}}{\textrm{M}_\odot/h} = 10^{12.19^{+0.06}_{-0.06}}
\left\{
\!
\begin{aligned}
\Bigg(\frac{L}{L^\star(z)}\Bigg)^{2.08^{+0.12}_{-0.12}} & \quad L<L^\star \\
\Bigg(\frac{L}{L^\star(z)}\Bigg)^{0.63^{+0.12}_{-0.12}} & \quad L>L^\star
\end{aligned}
\right.
\end{eqnarray}
where the slopes above and below $L^\star(z)$ are quite different. The best-fits show a steeply increasing halo mass with line luminosity up to $L^\star(z)$ with a slope of $2.08\pm0.12$ followed by a slowly increasing halo mass at brighter line luminosities with a slope of $0.63\pm0.12$ and a typical halo mass of  $10^{12.19\pm0.06}$ \msol~at $L^\star(z)$. 

\subsubsection{What causes the trend change at $L > L^\star(z)$?}
The slope change that is seen in Figure \ref{fig:lum} could be due to a change in the nature of the population of LAEs (e.g., \lya~emission is no longer driven by star formation but by AGN activity). This would result in the fraction of star-forming galaxies to decrease with increasing luminosity. Above $10^{12}$ \msol, the star formation efficiency decreases due to accelerated gas accretion caused by the deeper gravitational potentials of higher mass halos resulting in fewer star-forming galaxies with increasing halo mass (e.g., \citealt{Dekel2006,Bower2017}). This idea of a transitional or characteristic halo mass has been observed for \ha, \oiii, and \oii-selected emitters between $z \sim 0.4 - 5$ \citep{Sobral2010,Khostovan2018} and by studies of star-forming and passive galaxies (e.g., \citealt{Hartley2013,Dolley2014}).

To understand whether AGN contribution could be causing a trend change at $L > L^\star(z)$, we include the $z \sim 2 - 3$ AGN fraction measurements of \citet{Sobral2018_AGN} and $z \sim 1$ measurements of \citet{Wold2014} in the top panel of Figure \ref{fig:lum}. About 20 percent of $z\sim 1-3$ LAEs are found to be AGN around $L^\star$ and by $2 L^\star(z)$, half of the population of LAEs are AGNs. Calhau et al., submitted, found a strong correlation between the fraction of X-ray detected AGNs and \lya~luminosity. \citet{Matthee2017} found that $z \sim 2.3$ LAEs are about 50 percent X-ray AGNs at $> 10^{44}$ erg s$^{-1}$ (see also \citealt{Konno2016}). The halo masses measured for our $>2\ L^\star(z)$ samples are also consistent with previous AGN clustering studies (e.g., halo masses of $\gtrsim 10^{12.5}$ \msol; \citealt{Hickox2009,Koutoulidis2013,Allevato2016,Mendez2016,Hale2018}). Overall, we find that the brightest LAEs in our samples show properties that are consistent with AGN populations at high-$z$.

\subsection{Rest-Frame UV Continuum}
\label{sec:muv}

In the previous section, we found that the line luminosity properties of LAEs correlates with the host halo, regardless of redshift, such that the brightest LAEs reside in the most massive halos. Here we explore how the host halo mass can depend on the rest-frame UV properties, specifically the 1500\AA~UV continuum luminosity (\muv) and the UV-measured star formation rate. Our method of measuring both properties is described in \S\ref{sec:UV_prop}.

\begin{figure}
	\centering
	\includegraphics[width=\columnwidth]{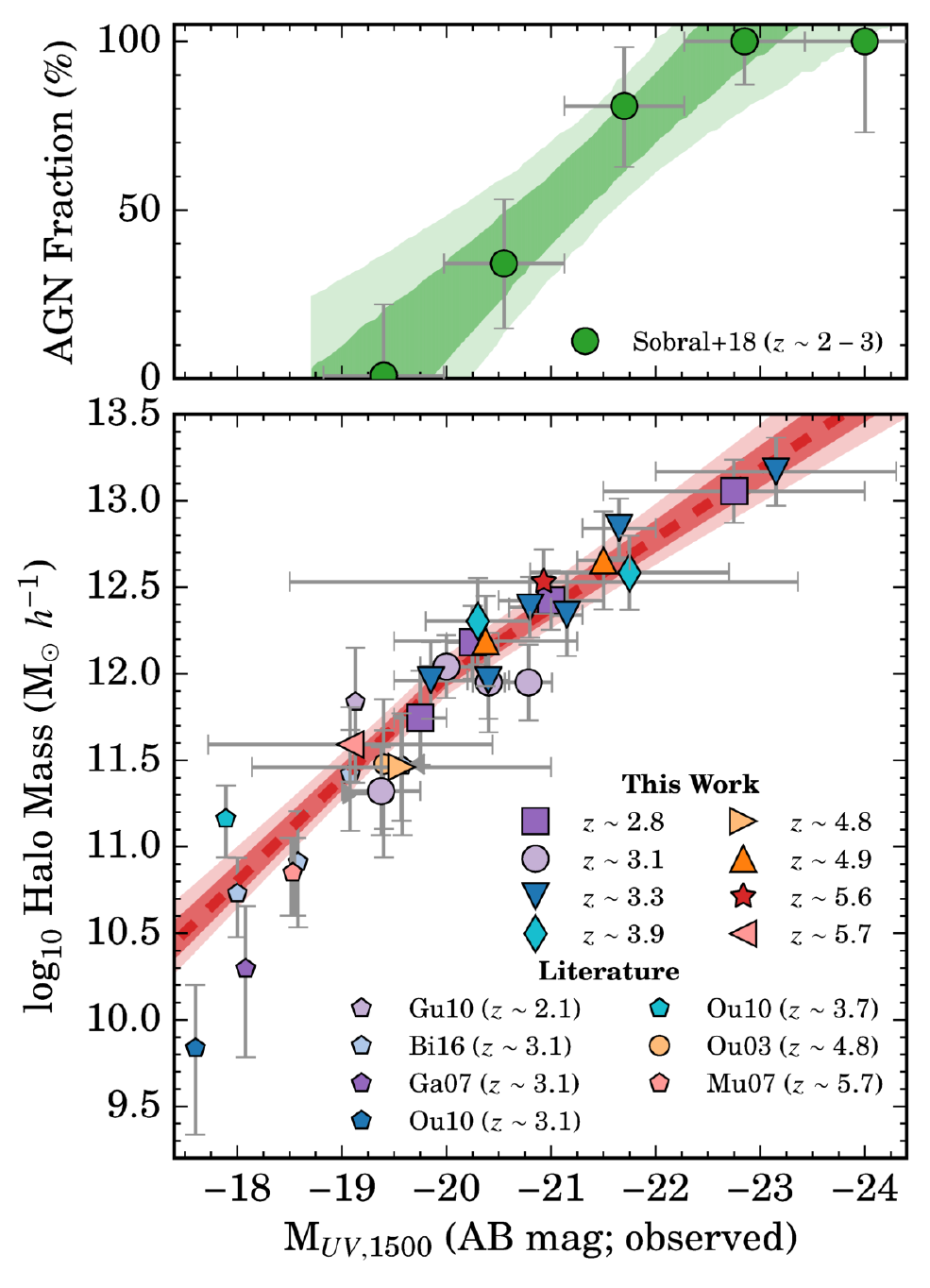}
	\caption{The host halo mass versus the rest-frame 1500 \AA~UV continuum luminosity for each redshift bin. We observe a strong, redshift-independent trend for the full range of UV luminosities observed where galaxies with the brightest continuum reside in massive halos. Our best-fit model is shown as a {\it red, dashed line} with the $1\sigma$ and $2\sigma$ regions highlighted as {\it dark} and {\it light red} regions, respectively. Included are the \muv~limit measurements from the literature. \citet{Bielby2016} covered multiple \muv~limits and also found a similar trend and even extend our observed trends down to $\textrm{\muv} \sim -18$. The AGN fraction measurements of \citet{Sobral2018_AGN} are shown in the {\it top panel} and show that our UV-bright LAEs are primarily AGNs.}
	\label{fig:muv}
\end{figure}

Figure \ref{fig:muv} shows how the observed (not corrected for dust) \muv~and the host halo mass are correlated. We find a strong trend where the host halo mass increases with increasing UV luminosity. The most UV-bright LAEs (\muv$ < -22$) are found to reside in $10^{13}$ \msol~halos and the fainter ones (\muv$ > -20$) are found in $< 10^{11.5}$ \msol~halos. We find a redshift-independent trend without the need to normalize by the characteristic UV luminosity, M$^\star_\textrm{UV}$. Interestingly, M$^\star_\textrm{UV}(z)$ has been observed by previous work to be constant within the redshift range of our samples (e.g., \citealt{Oesch2010,Alavi2016}). This shows that the redshift-independent trends are not an artifact of the model-dependent Schechter parameters.

We also include \muv-limit measurements from the literature which cover the faintest end of Figure \ref{fig:muv} \citep{Ouchi2003,Gawiser2007,Murayama2007,Guaita2010,Ouchi2010}. Presently, \citet{Bielby2016} is the only work that covered multiple \muv-limit thresholds for which they measured halo masses. Their measurements cover the range $-18 < \textrm{\muv} < -20$ and show an increasing trend between \muv~and halo mass in perfect agreement with the trends we observe with our samples. Furthermore, their \muv$ > -19$ measurements complement ours by showing that the trends seen at brighter UV luminosities continues down to $\textrm{\muv} \sim -18$. 

Using both our measurements and those from the literature, we fit a piecewise power law:
\begin{eqnarray}
\frac{\textrm{M}_\textrm{halo}}{\textrm{M}_\odot/h} = 
\left\{
\!
\begin{aligned}
10^{11.99^{+0.05}_{-0.06} - 0.40^{+0.03}_{-0.04} \Big(\textrm{\muv} + 20 \Big)} & \quad \textrm{\muv} > -20 \\
10^{11.99^{+0.05}_{-0.06} - 0.60^{+0.10}_{-0.13}\Big(\textrm{\muv} + 20 \Big)} & \quad \textrm{\muv} < -20
\end{aligned}
\right.
\end{eqnarray}
with the pivot at $\textrm{\muv} = -20$ mag, which is consistent with a changing slope towards fainter UV luminosities as shown by the literature measurements. The different slopes are statistically significant ($> 1\sigma$) and show a typical host halo mass of $\sim 10^{12}$ \msol~at $\textrm{\muv} \sim -20$ mag. This is similar to what we find for typical $L^\star(z)$ galaxies as shown in Figure \ref{fig:lum}, although the trend change is not as statistically significant. 

We also include the AGN fraction measurements of \citet{Sobral2018_AGN} in the top panel of Figure \ref{fig:muv}. The AGN fraction is found to increase with UV luminosity, such that 50 percent of LAEs are AGNs by $\textrm{\muv} \sim -21$ mag. We find halo masses of $>10^{12.5}$ \msol~within the AGN-dominated regime, which is consistent with other clustering studies of AGNs (e.g., \citealt{Hickox2009,Mendez2016}). This is also similar to what we find for the halo mass dependency with \lya~luminosity such that the brightest LAEs in terms of \lya~and UV luminosity at high-$z$ are consistent with AGN properties.

\subsection{Star Formation Rate}

\begin{figure}
	\centering
	\includegraphics[width=\columnwidth]{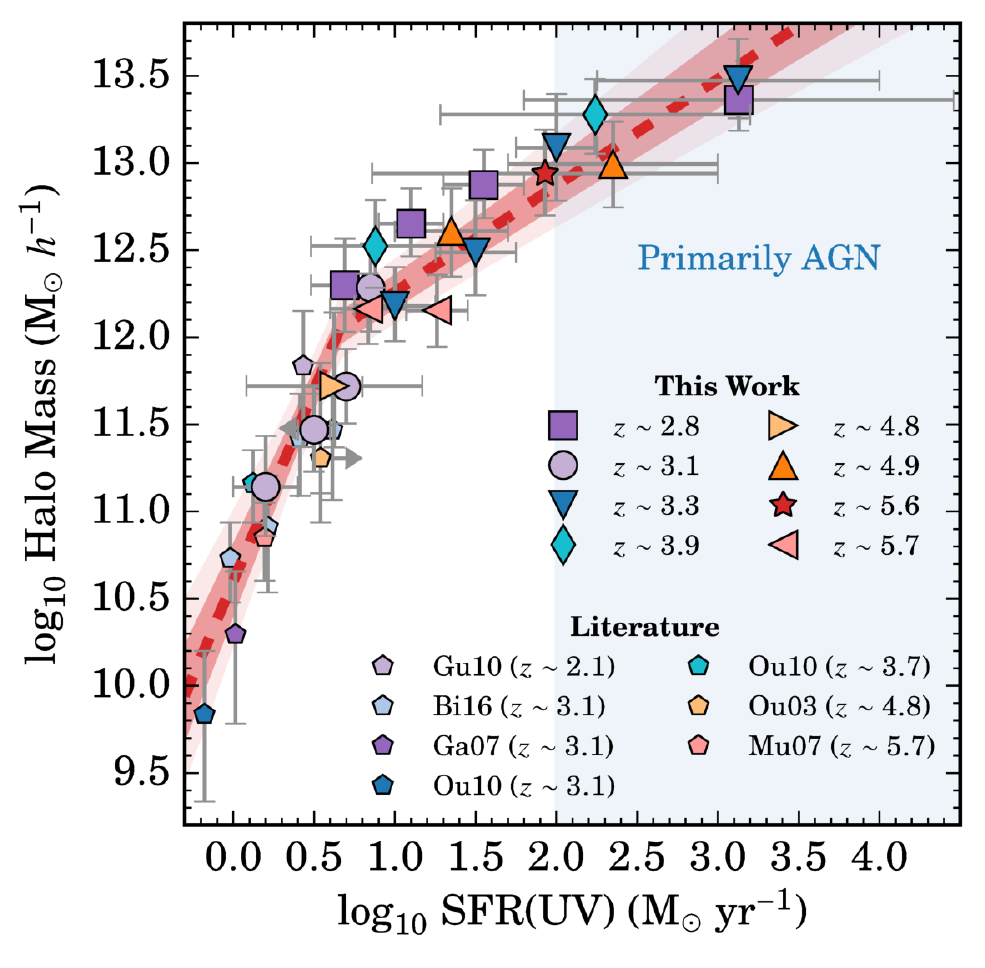}
	\caption{The host halo mass as measured in bins of dust-corrected rest-frame UV star formation rate. We find that an increasing, redshift-independent trend between increasing halo mass and increasing star formation rate. Our best-fit model is shown as a {\it red, dashed line} with the $1\sigma$ and $2\sigma$ regions highlighted as {\it dark} and {\it light red} regions, respectively. Included are the dust-corrected \muv-limit literature measurements from various narrowband surveys. \citet{Bielby2016} covers multiple star formation rate bins and also shows a similar trend in comparison to our observations, although for a limited star formation rate range. Above 10 \msol~yr$^{-1}$ and halo masses of $10^{12}$ \msol, the observed trends become shallower, similar to our observations of the halo mass - \lya~luminosity trends.}
	\label{fig:sfr_uv}
\end{figure}

The results in Figure \ref{fig:muv} are based on the observed \muv~for which the UV luminosity is not corrected for dust. To explore how host halo mass depends directly on the star formation rate, we dust correct \muv~using the UV slope, $\beta_{\rm UV}$, and use the \citet{Kennicutt1998} calibration as described in \S\ref{sec:UV_prop}. 

Figure \ref{fig:sfr_uv} shows the host halo mass for each LAE sample in bins of UV star formation rates. We find that the host halo mass increases with increasing star formation rate at all redshifts. The trends observed are also found to be redshift independent, similar to the other trends with galaxy properties that have been noted in this paper. The range of dark matter halo masses shown vary greatly with the least active galaxies ($\textrm{SFR} \sim 1.6$ \msol~yr$^{-1}$) residing in $10^{11.2}$ \msol~halos and the most active ($\textrm{SFR} \sim 100$ \msol~yr$^{-1}$) residing in $10^{13}$ \msol~halos. SFRs $> 100$ \msol~yr$^{-1}$ primarily have their UV continuum emission powered by AGNs as we saw in Figure \ref{fig:muv} and, therefore, should be interpreted with caution. This region is highlighted in Figure \ref{fig:sfr_uv}.
	
Included in Figure \ref{fig:sfr_uv} are the \muv-limit measurements from $z\sim 2 - 6$ studies found in the literature \citep{Ouchi2003,Gawiser2007,Murayama2007,Guaita2010,Ouchi2010,Bielby2016}. The measurements were redone to match with the assumptions made in this work (see Appendix \ref{sec:literature} for details) and converted to SFR using Equation \ref{eqn:sfr}. The typical $\beta_{\rm UV}$ slope for these samples are bluer than $\beta_{\rm UV} \sim -2$, which implies zero to minimal dust attenuation (e.g., see Figure 2 of \citealt{Ono2010}). Because these measurements are \muv~(SFR)-limit studies, they help to constrain the least active end ($\textrm{SFR} \lesssim 1.6$ \msol~ yr$^{-1}$) of Figure \ref{fig:sfr_uv}.

We find that two trends are present in Figure \ref{fig:sfr_uv} where the halo mass increases rapidly from low SFR to $\sim 4.5$ \msol~yr$^{-1}$ and continues to increase with a shallower slope to higher SFRs. To quantify these trends, we fit our measurements and those from the literature with a piecewise power law. The best fit is:
\begin{eqnarray}
\frac{\textrm{M}_\textrm{halo}}{\textrm{M}_\odot/h} = 10^{12.05^{+0.08}_{-0.09}}
\left\{
\!
\begin{aligned}
\Bigg(\frac{\textrm{SFR}}{4.5}\Bigg)^{2.19^{+0.25}_{-0.23}} & \quad \textrm{SFR} < 4.5~ \frac{\textrm{\msol}}{\textrm{yr}} \\
\Bigg(\frac{\textrm{SFR}}{4.5}\Bigg)^{0.61^{+0.09}_{-0.05}} & \quad \textrm{SFR} > 4.5~ \frac{\textrm{\msol}}{\textrm{yr}}
\end{aligned}
\right.
\end{eqnarray}
with a typical halo mass of $10^{12.05^{+0.08}_{-0.09}}$ \msol~at $\textrm{SFR} \sim 4.5$ \msol~yr$^{-1}$, which is the point for which we visually see a change in the trend in Figure \ref{fig:sfr_uv}. 

In comparison to the halo mass - $L_\textrm{\lya}$ trend we measured, there are many important similarities. The pivot point in the piecewise has similar halo masses and the slopes of both trends are very much similar. This could suggest that $L_\textrm{\lya}$ is indeed tracing the star formation activity, despite the many caveats surrounding using \lya~as a star formation indicator (see \citealt{Dijkstra2017} for a review). 

The typical halo mass measured at SFR$ = 4.5$ \msol~yr$^{-1}$ is consistent with the peak of star formation efficiency found in halos of $\sim 10^{12}$ \msol. This is similar to what we also find for the halo mass - $L_\textrm{\lya}$ results. The changing slope seen above $4.5$ \msol~yr$^{-1}$ is most likely due to the combined effects of a larger population of AGN and the existence of LAEs that are undergoing an intense period of star formation activity. 

The observed trends suggest that the processes that govern star formation activity and the production of the \lya~line in LAEs are strongly tied to the host halo mass properties. The redshift independence reinforces the idea that this connection is independent of time such that halos and their residing galaxies co-evolve with each other in unison. This would then suggest that one of the most important characteristics that governs the evolution of a LAE is the host dark matter halo mass.

\subsubsection{Comparisons to UV-selected Samples}

\begin{figure}
	\centering
	\includegraphics[width=\columnwidth]{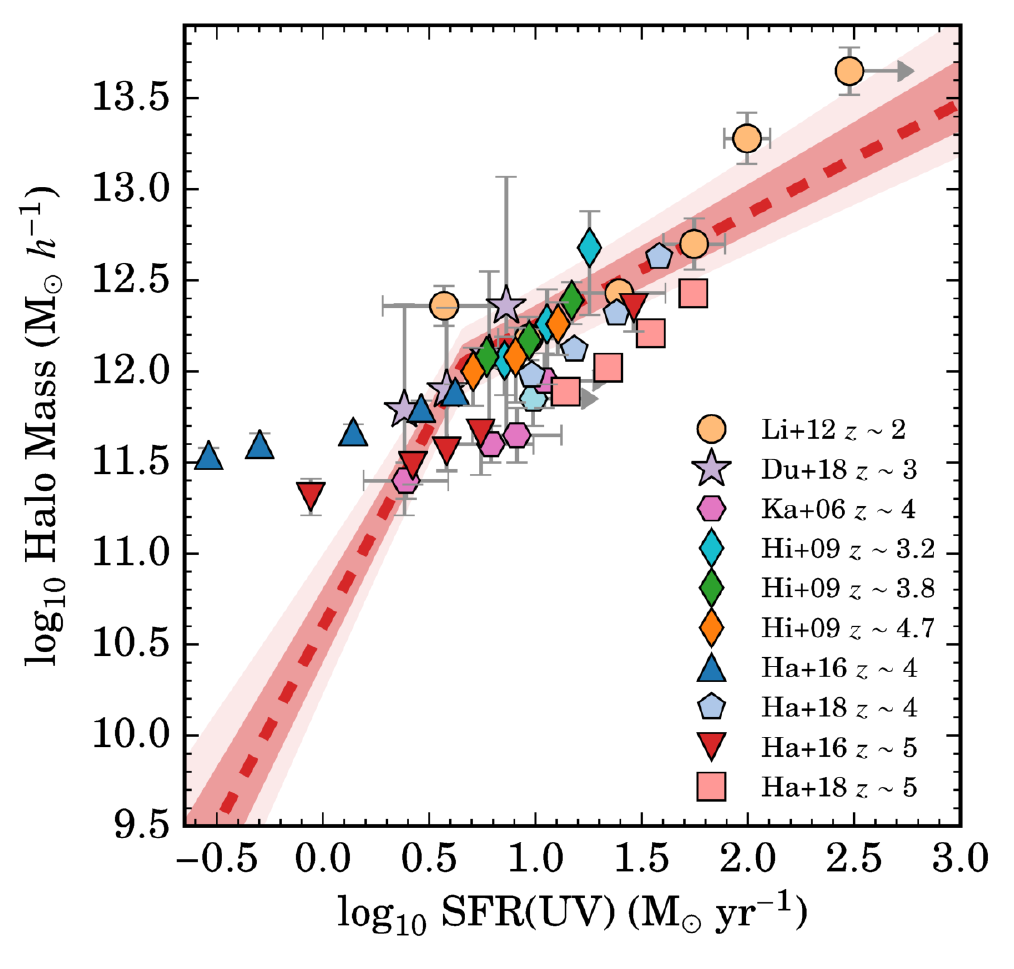}
	\caption{Comparison between our best-fit halo mass -- UV star formation rate relation for LAEs (shown as a {\it dashed red line}) and a compilation of measurements from $z \sim 2 - 5$ UV-selected samples: $BzK$ \citep{Lin2012}, LBGs \citep{Kashikawa2006,Hildebrandt2009,Harikane2016,Harikane2018}, and the VUDS spectroscopic survey \citep{Durkalec2018}. The $1\sigma$ and $2\sigma$ regions of our model are highlighted in {\it dark} and {\it light red}, respectively. The literature measurements are in agreement with our model for SFR $\gtrsim 3$ \msol~yr$^{-1}$ ($\log_{10}$ SFR $\gtrsim 0.5$), although with some scatter which can be due to the different methods used in measuring halo masses in each study (e.g., different clustering slopes, halo occupation model prescriptions, halo mass and bias models, wide redshift distributions). Below 3 \msol~yr$^{-1}$, LBGs are measured to reside in higher mass halos in comparison to LAEs.}
	\label{fig:SFR_lit}
\end{figure}

Throughout this paper we have focused on LAEs and how they relate to their host halo properties. Here, we investigate how our results relate to UV-selected samples. We use a compilation of halo mass measurements from the literature that are selected as $z\sim2$ $BzK$ \citep{Lin2012} and $z \sim 2 - 5$ LBGs \citep{Kashikawa2006,Hildebrandt2009,Harikane2016, Harikane2018}. We also include the recent $z\sim3$ VUDS spectroscopic survey measurements \citep{Durkalec2018}.

Figure \ref{fig:SFR_lit} shows the comparison of our results with UV-selected samples drawn from the literature. All measurements from the literature confirm an increasing halo mass with increasing star formation rate. We find that the $z\sim2$ $BzK$ measurement of \citet{Lin2012} is in agreement with our measurements to star formation rates of $\sim 300$ \msol~yr$^{-1}$ (corrected for the different calibration used in their study). The VUDS measurements of \citet{Durkalec2018} are also in agreement down to $\sim 3$ \msol~yr$^{-1}$. 

The majority of literature measurements shown in Figure \ref{fig:SFR_lit} are from LBG-selected samples at $z \sim 2 - 5$. We find our measurements are in agreement with LBG studies at SFR $\gtrsim 3$ \msol~yr$^{-1}$, while a deviation is seen at lower SFRs. The typical LBG at $z\sim 4 - 5$ with SFR $\lesssim 3$ \msol~yr$^{-1}$ is found to reside in $10^{11.3 - 11.7}$ \msol~halos, while we find LAEs reside in significantly lower mass halos with decreasing star formation rate in respect to LBGs. Our result is in agreement with \citet{Bielby2016}, where they measured the LAE-LBG cross-correlation function at $z\sim3$ and concluded that LAEs comprise the low-luminosity portion of LBGs that reside in low-mass halos. 

We note that our results and those from the literature shown in Figure \ref{fig:SFR_lit} are not entirely compatible for a direct comparison. This is due to the different methods used in measuring halo mass ranging from the different clustering slopes and methodology in quantifying the correlation functions (e.g., incorporating redshift distributions, Limber approximation) to the halo model prescriptions used (e.g., halo occupation distributions, halo mass and bias functions). All these points need to be taken into account for a proper quantitative comparison, although, qualitatively, we find that LAEs only seem to deviate from the halo properties of UV-selected samples at lower UV star formation rates.

\begin{figure}
	\centering
	\includegraphics[width=\columnwidth]{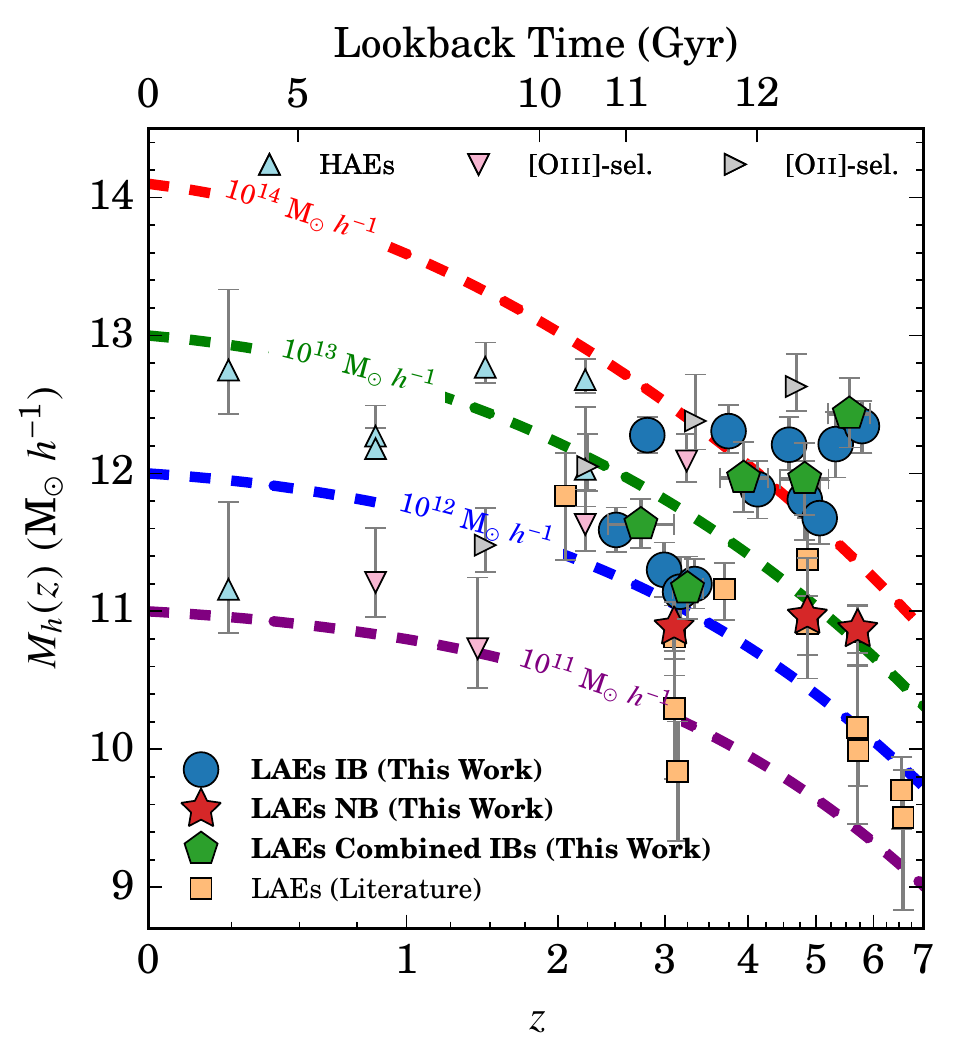}
	\caption{The present-day descendants of emission line-selected galaxies. The {\it dashed lines} are the evolutionary tracks of present-day $10^{11-14}$ \msol~host halos assuming the halo mass accretion model of \citet{Behroozi2013}. Our narrowband-selected LAEs are found to be progenitors of galaxies in present-day $10^{12 - 13}$ \msol~halos (`Milky Way-like'), while our intermediate band samples are in more massive present-day halos ($> 10^{13}$ \msol~halos; 'cluster-like'). Other LAE narrowband surveys show a similar result such that \lya~emiters are the progenitors of a wide range of present-day galaxies, similar to what is found for \ha-, \oii-, and \oiii-selected galaxies.}
	\label{fig:descendants}
\end{figure}

\subsection{Present-Day Descendants of LAEs}

At all redshift slices, we find that LAEs cover a wide range in host halo mass and redshift-independent trends between halo mass and various galaxy properties. But what types of galaxies do LAEs evolve into in respect to the present-day? We can address this question by using halo mass accretion models to predict the present-day halo mass of a host halo at a given redshift. 

Figure \ref{fig:descendants} shows the present-day halo masses of our full LAE samples measured using the halo mass accretion tracks of \citet{Behroozi2013}. Given the wide range in halo masses we have observed throughout this study, we find that LAEs cover a wide range of present-day descendants from dwarf-like ($M_\textrm{halo}(z=0) \sim 10^{11}$ \msol), to Milky Way-like ($M_\textrm{halo}(z=0)  \sim 10^{12}$ \msol), and galaxies residing in cluster-like environments ($M_\textrm{halo}(z=0) > 10^{13}$ \msol) all the way to the richest clusters. Since we find the brightest LAEs are highly clustered and reside in $10^{13}$ \msol~halos, they are most likely the `protoclusters' that are known to be the densest regions in the high-$z$ Universe and evolve to become the massive, rich galaxy clusters we see in the local Universe (e.g., \citealt{Franck2016,Overzier2016} and references therein). Our results suggest that bright LAEs are useful candidates in selecting potential overdense regions for further study on massive cluster formation.

The wide range in the type of descendants shows that LAEs are great tools in studying how galaxies formed and evolved to the ones we currently see in the local Universe. Other narrowband-selected samples overlaid in Figure \ref{fig:descendants} also show a wide range in present-day descendants, suggesting that narrowband surveys, in general, can provide us with samples of star-forming galaxies at various redshifts to map out the evolutionary track of galaxies from high-$z$ to the present-day.

\section{Conclusions}
\label{sec:conclusion}

We present a comprehensive investigation of the clustering and halo properties of $\sim 5000$ LAEs and explore their dependences on \lya~luminosity, UV continuum luminosity, and UV star formation rate in multiple redshift slices between $z \sim 2.5 - 6$. We highlight the main results of this study here:

\begin{enumerate}[leftmargin=*]
	\item The clustering lengths of the narrowband LAE samples are shown to increase from $\textrm{\ro} \sim 3.1 - 5.0$ Mpc $h^{-1}$ between $z \sim 3.1 - 5.7$. The intermediate band and combined intermediate band samples show a more rapid increase in \ro~from $\sim 4.5$ Mpc $h^{-1}$ at $z \sim 2.5$ to $\sim 16$ Mpc $h^{-1}$ by $z \sim 5.8$. The typical halo masses of the narrowband samples are found to be $\sim 10^{11}$ \msol, while the intermediate band samples range between $\sim 10^{11 - 12}$ \msol.	
	
	\item Host halo mass is found to increase with increasing \lya~luminosity at all redshifts probed. 
	 
	\item Normalizing \lya~luminosity by $L^\star(z)$ shows a redshift-independent trend with host halo mass. LAEs are found to reside in a wide range of host halos ranging from $10^{9.75}$ \msol~at $\sim 0.1 L^\star(z)$ to $10^{12.1}$ \msol~at $L^\star(z)$ and $10^{12.8}$ \msol~at $\sim 10 L^\star(z)$. 
	
	\item We find a strong, redshift-independent trend between host halo mass and observed 1500\AA~UV continuum luminosity. LAEs with $\textrm{\muv}\sim -18$ mag are found to reside in $10^{10.5}$ \msol~halos and $\textrm{\muv} \sim -23$ mag in $10^{13}$ \msol~halos. 
	
	\item We also find a strong, redshift-independent trend between host halo mass and dust-corrected UV star formation rate. We find that LAEs with SFR $\sim 1$ \msol~yr$^{-1}$ reside in $10^{10}$ \msol~halos and $\sim 100$ \msol~yr$^{-1}$ reside in $10^{12.8}$ \msol~halos.
	
	\item For both \lya~luminosity and UV SFR, we observe sharp trend changes. In the case of \lya~luminosity, we find that the host halo mass scales as $(L/L^\star(z))^{2.08}$ and $(L/L^\star(z))^{0.63}$ for below and above $L^\star(z)$, respectively. A similar trend is seen between halo mass and SFR with the trend change occurring at $\sim 4.5$ \msol~yr$^{-1}$. This is attributed to a changing population of LAEs where the brightest LAEs, in terms of line luminosity and SFRs, primarily have their emission powered by AGNs and not star formation activity. 
	
	\item We find that \lya~emitters are progenitors of a wide range of present-day galaxies depending on their \lya~luminosity ranging from dwarf-like systems (similar to LMC and SMC) to galaxies in cluster-like environments, such that \lya~emitters are great tools to understand the evolutionary path of galaxies from high-$z$ to the local Universe.
		
\end{enumerate}

Our results highlight the significant connection that host halos and galaxies share from the end of the epoch of reionization to the peak of cosmic star formation. The redshift-independent trends with halo mass signify the co-evolution of galaxies and their host halos and emphasis the importance of halos in the overall evolution of galaxies. The results presented in this paper provide empirical relations that can be tested in simulations and provide new constraints in regards to the physics behind galaxy evolution. Our results also emphasize the importance of investigating clustering and halo properties in terms of various galaxy properties to take selection effects into account when comparing samples from different surveys.

\section*{Acknowledgments}
AAK acknowledges that this work was supported by NASA Headquarters under the NASA Earth and Space Science Fellowship Program - Grant NNX16AO92H. JM acknowledges support from the ETH Zwicky fellowship. RKC acknowledges funding from STFC via a studentship. APA acknowledges support from the Funda\c{c}\~ao para a Ci\^encia e a Tecnologia FCT through the fellowship PD/BD/52706/2014 and the research grant UID/FIS/04434/2013. JC and SS both acknowledge their support from the Lancaster University PhD Fellowship.

We have benefited greatly from the publicly available programming language Python, including the {\sc NumPy}, {\sc SciPy}, {\sc Matplotlib}, {\sc scikit-learn}, and {\sc astropy} packages, as well as the {\sc TOPCAT} analysis program.

The SC4K samples used in this paper are all publicly available for use by the community \citep{Sobral2018}. The catalog is also available on the COSMOS IPAC website (\url{https://irsa.ipac.caltech.edu/data/COSMOS/overview.html}).

\bibliography{clustering_LyA}

\appendix

\section{Effects of the SA22 Overdense Region}
\label{sec:overdense} 
 
\begin{figure}
	\centering
	\includegraphics[width=\columnwidth]{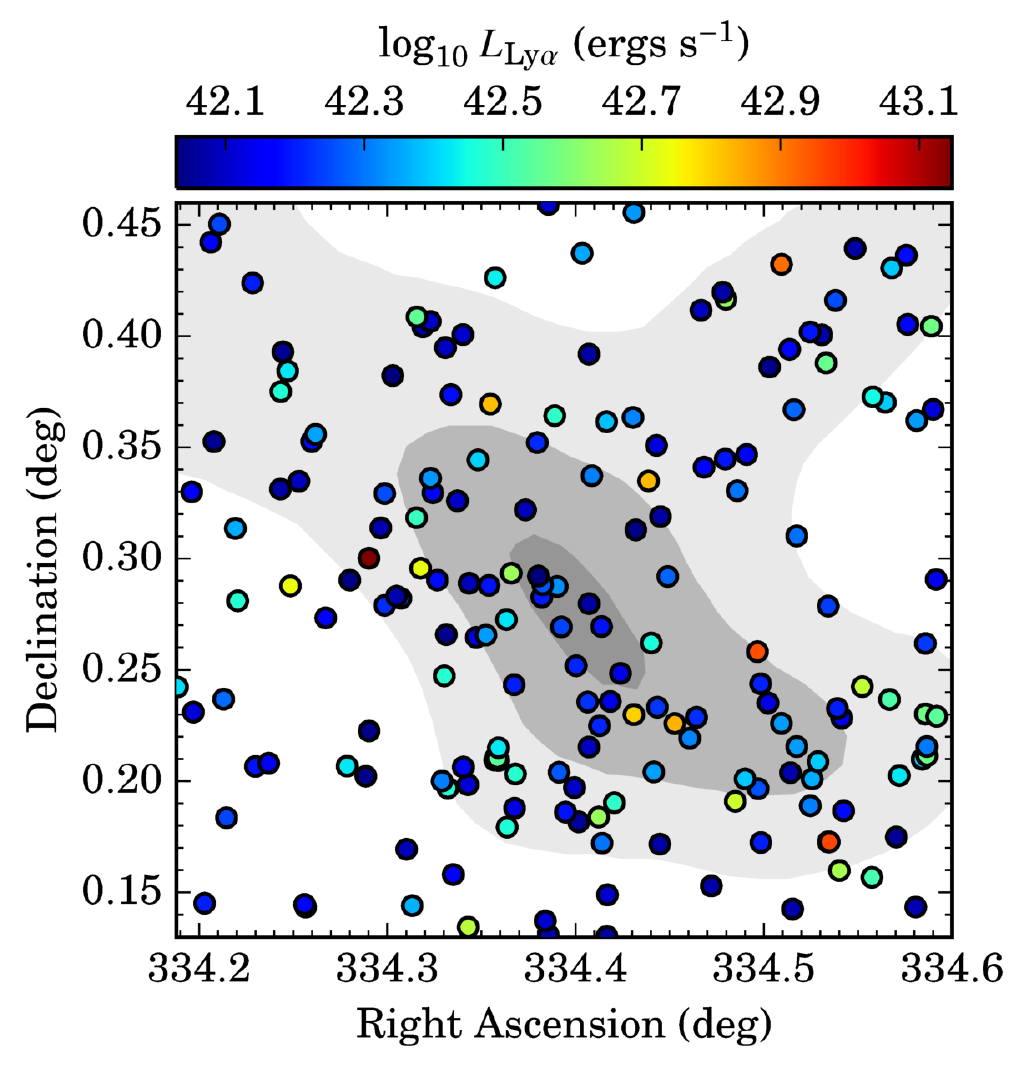}
	\caption{A $24' \times 18'$ cutout of the SA22 field centered on the $z\sim3.1$ protocluster. The {\it circles} represent the angular positions of our NB497-selected LAEs with the color scale related to their line luminosities. The {\it grey shaded regions} are the source isodensity levels from \citet{Hayashino2004} and compiled by \citet{Saez2015} that highlight the location of the SA22 overdense region. The $z=3.1$ LAEs that seem to primarily populate the overdense region ({\it darker grey regions}) have line luminosities of $\lesssim 10^{42.4}$ erg s$^{-1}$, which are the faintest LAEs in the sample.}
	\label{fig:NB497_overdensity}
\end{figure}

Previous work on the SA22 field has identified and extensively studied a significant overdense region (protocluster) comprised of \lya~emitters and LBGs at $ z\sim 3.1$ (e.g., \citealt{Steidel1998,Steidel2000,Hayashino2004,Webb2009,Nestor2011,Yamada2012,Kubo2015,Kubo2016,Saez2015,Topping2016}). Figure \ref{fig:NB497_overdensity} shows a $0.12$ deg$^{2}$ cutout centered on the position of the overdense region highlighted by the source isodensity levels from \citet{Hayashino2004}. We include the distribution of our $z \sim 3.1$ LAEs and find that it traces the underlying overdensity. The main question that needs to be addressed is how does the overdense region affect our observed angular correlation functions as shown in the top panel of Figure \ref{fig:acf_NB}?

As described in \S\ref{sec:acf}, there is a signature of a $1-$halo term in the NB497 ACFs which would suggest a satellite fraction of LAEs. We investigate if this is due to the presence of the significant overdense region discussed above by measuring the ACFs using the same methodology used throughout this paper and masking the $0.12$ deg$^2$ field shown in Figure \ref{fig:NB497_overdensity}. Note that we also mask the random maps as well to ensure a consistent survey geometry.

\begin{figure}
	\centering
	\includegraphics[width=\columnwidth]{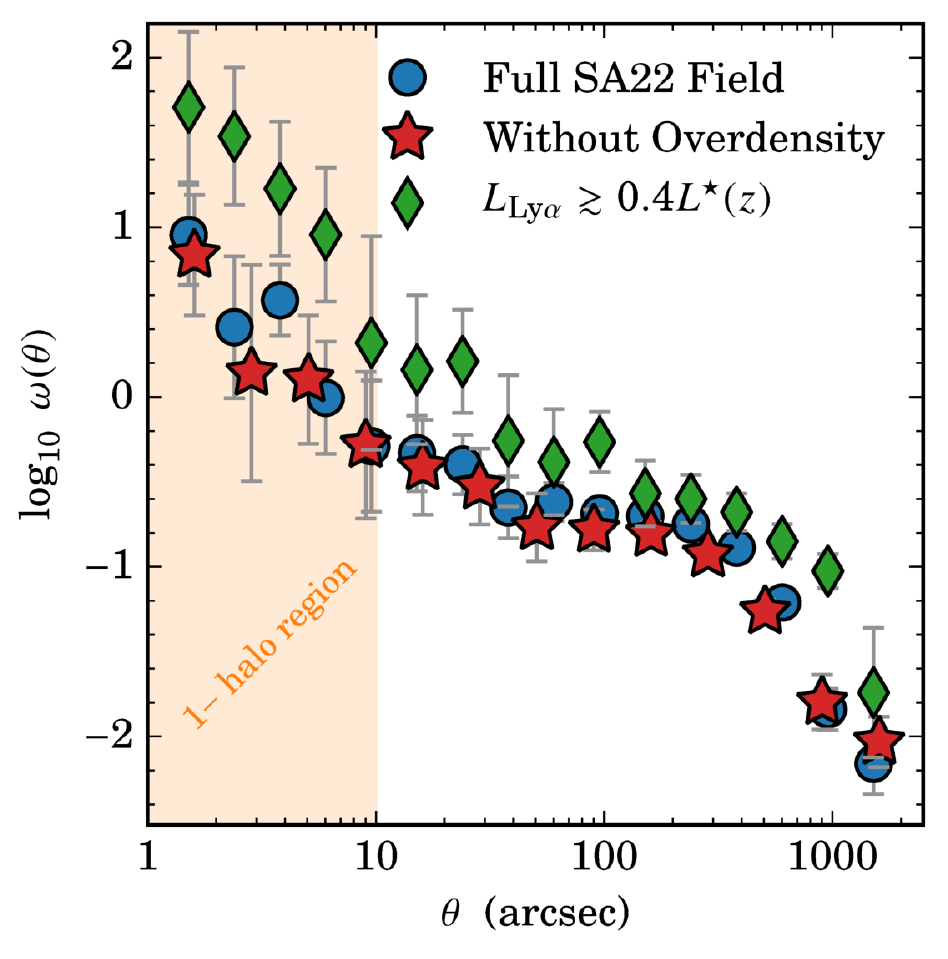}
	\caption{Comparison of the ACF for the Full SA22 field ({\it blue circles}) and for the case where the overdensity is masked ({\it red stars}). We find that both ACFs are consistent with each other, such that the $z \sim 3.1$ protocluster is not affecting our clustering measurements and that the $1-$halo term still exists in both cases (region highlighted in {\it orange}). The {\it green diamonds} show the ACF for the Full SA22 field with the added condition of a line luminosity threshold $\gtrsim 0.4 L^\star$. We find that above this threshold, the deviation from a simple power law associated with the detection of a $1-$halo term disappears, suggesting that the satellite fraction of galaxies is close to zero above $0.4 L^\star$.}
	\label{fig:NB497_ACF_overdense}
\end{figure}

Figure \ref{fig:NB497_ACF_overdense} shows the comparison between the ACFs for the full field and for the case where the overdense region is masked. We find that there is no difference between the two correlation functions, such that the overdensity does not significantly affect the overall clustering measurements. \citet{Hayashino2004} also found the overdense region to have a weak clustering signal relative to the whole SA22 field.

We next investigate if the cause of the $1-$halo term is due to flux depth of the sample. Figure \ref{fig:NB497_overdensity} shows the spatial and luminosity distribution of the $z\sim3.1$ LAEs where concentrations of faint LAEs are found within the overdense region. To test if the $1-$halo term arises from faint LAEs that comprise the satellite population, we measure the ACFs for the full SA22 field with varying line luminosity thresholds. Figure \ref{fig:NB497_ACF_overdense} shows the $1-$halo term disappears by $L_\textrm{\lya} \gtrsim 0.4\ L^\star$, which is consistent with the spatial/luminosity distribution shown in Figure \ref{fig:NB497_overdensity}.

Although a detailed analysis of the satellite fractions is beyond the scope of this work, we find that all our samples do not exhibit signatures of population of satellite galaxies except for the NB497 sample, which we find to also have negligible satellite fraction effects above $0.4 L^\star$. If the satellite fractions do affect our measurements, then it would be primarily isolated to the lowest line luminosity ranges of our samples and would result in overestimations of their host dark matter halos. Lastly, the most comprehensive study of satellite fractions of emission line-selected galaxies was done by \citet{Cochrane2017} for their $z = 0.4$, $0.8$, $1.5$, and $2.2$ NB-selected \ha~samples where they find that, on average, $3 - 5$ percent of the sample are satellites. In comparison to LBG samples, the recent 100 deg$^2$ HSC survey found satellite fractions of $\lesssim 5$ percent between $z \sim 4 - 6$ \citep{Harikane2018}. Throughout the rest of this paper, we neglect the effects of the satellite population.

\section{Clustering Measurements from the Literature}
\label{sec:literature}

We have used several narrowband studies for the purpose of comparison throughout this paper. Due to our unique approach and the varying assumptions between each measurement drawn from the literature, we have to be careful about how we are comparing our measurements to the literature. To resolve this issue, we have to normalize the methodology between our clustering measurements and those from the literature. We achieve this by taking the observed angular correlation functions from each narrowband study, fit Equation \ref{eqn:scf} to measure \ro~and use the narrowband filter attributed to that study as the proxy for the redshift distribution, and include the errors associated with cosmic variance by using the empirical relation measured by \citet{Sobral2010}.

\begin{table*}
	\centering
	\caption{Clustering Measurements of \lya~Emitters from the Literature. Shown are the referred narrowband studies, redshifts per each sample, the corresponding narrowband filter identification, the survey area, the \ro~reported in the respective study, our measurement of \ro~based on our assumptions and methodology using the observed angular correlation functions, and the host halo mass using our own \ro-halo mass model.}
	\label{table:lit_r0}
	\begin{tabular*}{\textwidth}{l @{\extracolsep{\fill}}cccccc}
		\hline
		Study                & Redshift      & Filter &   Area    & \ro$_\textrm{,reported}$  & \ro$_\textrm{,measured}$  &  $\log_{10}$ Halo Mass  \\
		                     &               &        & (deg$^2$) &     (Mpc $h^{-1}$)   &     (Mpc $h^{-1}$)   &    (\msol~$h^{-1}$)     \\ \hline
		\citet{Hao2018}			  & $2.23\pm0.03$ & NB393  &   0.34    &
		$2.56\pm0.33$ 		& $3.02\pm0.53$ & $11.32^{+0.29}_{-0.40}$ \\
		\citet{Bielby2016}        & $3.10\pm0.03$ & NB497  &   1.07    & $2.86\pm0.33$    	& $2.99\pm0.40$ & $10.81^{+0.23}_{-0.28}$ \\
		\citet{Gawiser2007}    & $3.11\pm0.02$ & NB4990 &   0.27    & $2.52^{+0.56}_{-0.70}$ & $2.34^{+0.43}_{-0.43}$ & $10.30^{+0.36}_{-0.51}$ \\
		\citet{Guaita2010}       & $2.07\pm0.02$ & NB3727 &   0.36   & $4.80^{+0.90}_{-0.90}$ & $4.33^{+1.01}_{-1.01}$ & $11.84^{+0.31}_{-0.46}$ \\
		\citet{Murayama2007} & $5.71\pm0.04$ & NB816  &   1.95    & --- & $4.96^{+0.71}_{-0.71}$ & $10.85^{+0.20}_{-0.25}$ \\
		\citet{Ouchi2003}        & $4.86\pm0.03$ & NB711  &   0.15    & --- & $6.03^{+1.49}_{-1.49}$ & $11.37^{+0.30}_{-0.43}$ \\
		\citet{Ouchi2010}        & $3.14\pm0.03$ & NB503  &   0.98    & $1.70^{+0.39}_{-0.46}$ & $1.96^{+0.30}_{-0.30}$ & $9.84^{+0.50}_{-0.36}$ \\
		\citet{Ouchi2010}        & $3.69\pm0.03$ & NB570  &   0.96    & $2.74^{+0.58}_{-0.72}$ & $4.11^{+0.52}_{-0.52}$ & $11.16^{+0.19}_{-0.22}$ \\
		\citet{Ouchi2010}        & $5.71\pm0.04$ & NB816  &   1.03    & $3.12^{+0.33}_{-0.36}$ & $3.29^{+0.99}_{-0.99}$ & $10.16^{+0.45}_{-0.70}$ \\
		\citet{Ouchi2010}        & $6.55\pm0.05$ & NB921  &   0.90    & $2.31^{+0.65}_{-0.85}$ & $2.93^{+0.39}_{-0.39}$ & $9.70^{+0.24}_{-0.28}$ \\
		\citet{Ouchi2018}        & $5.72\pm0.05$ & NB816  &   13.8    & $3.01^{+0.35}_{-0.35}$ & $3.01^{+0.37}_{-0.37}$ & $9.99^{+0.22}_{-0.25}$ \\
		\citet{Ouchi2018}        & $6.58\pm0.05$ & NB921  &   21.2    & $2.66^{+0.49}_{-0.70}$ & $2.66^{+0.50}_{-0.71}$ & $9.50^{+0.34}_{-0.67}$ \\ 
		\citet{Shioya2009}		& $4.86\pm0.03$ & NB711  &   1.83    & $4.40^{+1.30}_{-1.50}$ & $4.44^{+0.59}_{-0.59}$ & $10.91^{+0.20}_{-0.23}$\\ \hline
	\end{tabular*}
\end{table*}

Table \ref{table:lit_r0} shows our recalculations of \ro~for each narrowband study used in this paper for comparison purposes. We also include the measured halo masses in Table \ref{table:lit_r0}, which are based on the same assumptions described in \S\ref{sec:dmh}. In comparison to the measurements reported in each study, the error bars we measure are typically larger than that reported in the literature due to the inclusion of cosmic variance effects.

We detail on a few of these studies as we had to apply specific corrections/extensions. For the \citet{Ouchi2003} study, only a measurement of the clustering amplitude and slope was reported for which we have extended this work by making measurements of \ro~and halo mass. The \citet{Murayama2007} angular correlation functions did not include an integral constraint correction. Since the survey size and sample is essentially the same as our NB816 sample (note that we used the archival NB816 images which are the same used in \citealt{Murayama2007}), we use our integral constraint to correct their angular correlation functions. For the \citet{Shioya2009}, we report their measurement of $4.4^{+1.3}_{-1.5}$ Mpc $h^{-1}$ which assumes a slope of $\gamma = -1.90 \pm 0.22$. In our recalculation of their measurement, we keep $\gamma$ fixed on $-1.80$, which is consistent with their measured slope.

\section{Intermediate Band Angular Correlation Functions}

\begin{figure*}
	\centering
	\includegraphics[width=2\columnwidth,trim= 0 9 0 9,clip]{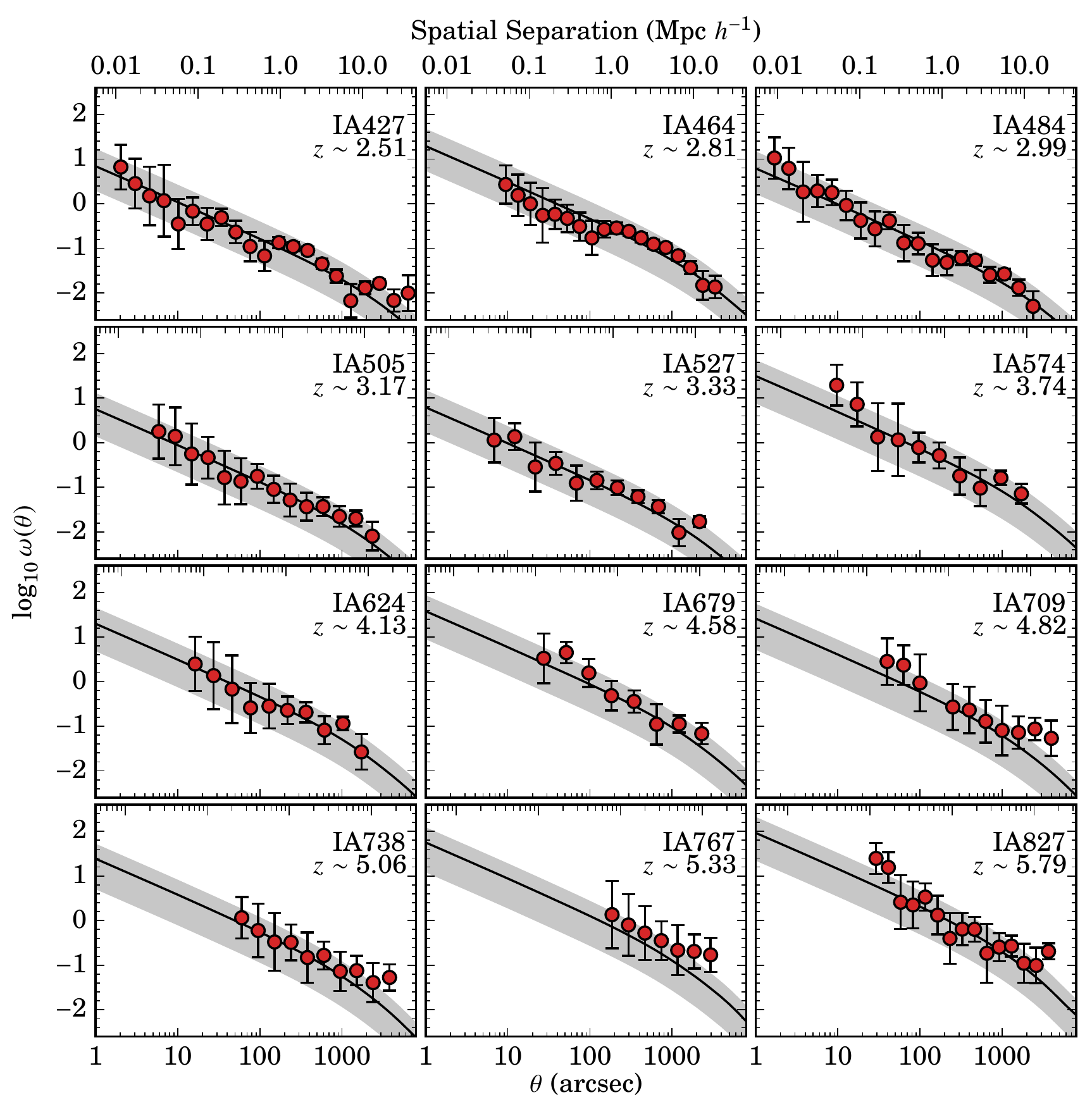}
	\caption{The angular correlation functions for each intermediate band sample. The {\it red circles} are the median observed measurements for \wtheta~based on all 2000 iterations. The {\it black line} shows the best-fit model as described in Equation \ref{eqn:scf} with the $1\sigma$ uncertainty represented as the {\it grey}. The spatial axis shown in each panel corresponds to the spatial separation for a given angular separation at the redshift of the samples shown. We find no significant detections of the 1-halo term, implying that satellite fractions are negligible for these LAEs. Note that the IB samples are biased towards the brightest LAEs (typically $L > L^\star(z)$ galaxies).}
	\label{fig:acf_IB}
\end{figure*}

\begin{figure*}
	\centering
	\includegraphics[width=2\columnwidth]{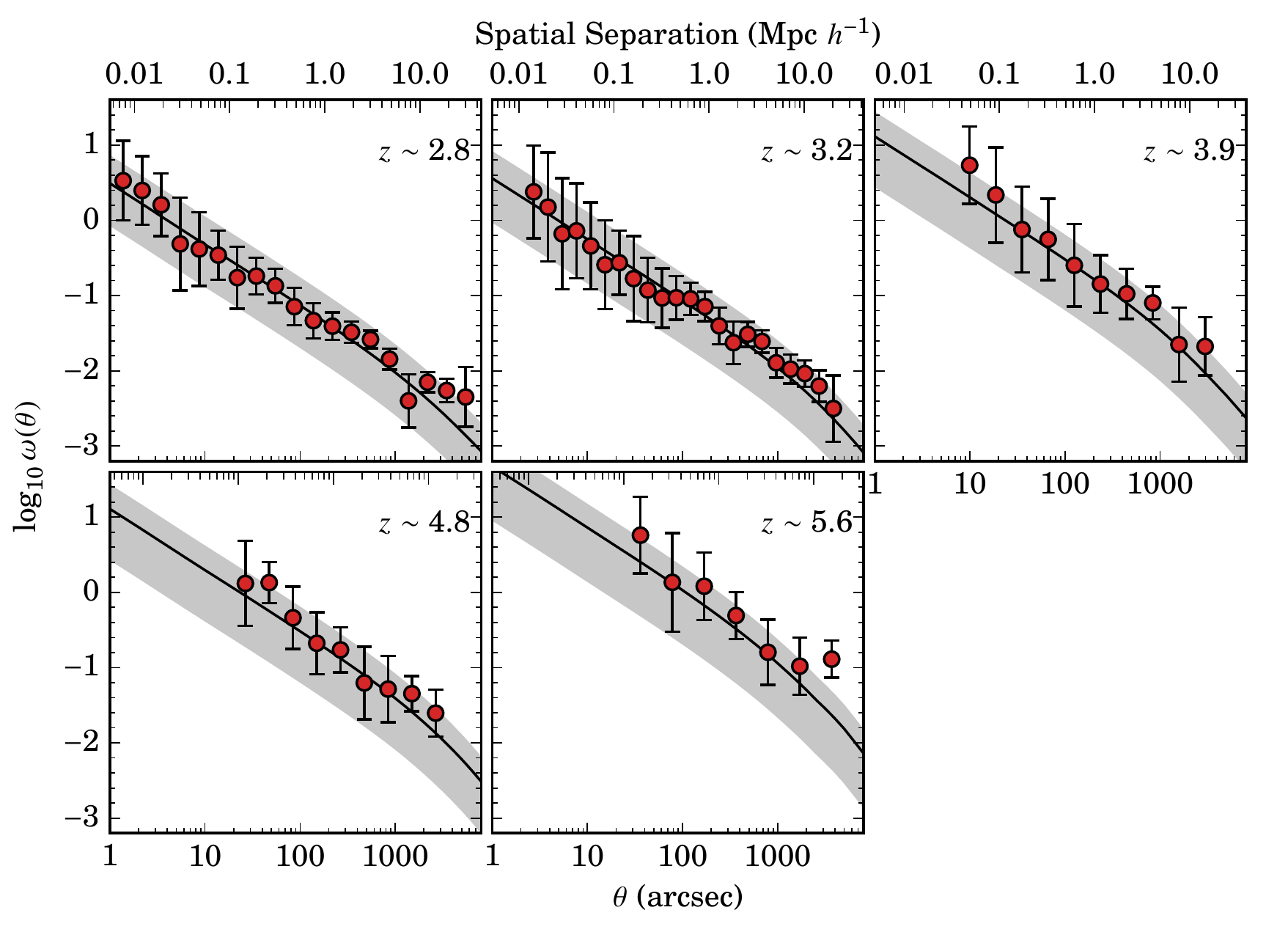}
	\caption{Same description as in Figure \ref{fig:acf_IB} but for the combined intermediate samples. The main importance of the combined IB samples is the large increase in sample sizes, especially at $z > 4$ where the individual IB samples contain $\sim 30 - 100$ LAEs each. As found in Figure \ref{fig:acf_IB}, we find no significant detection of a $1-$halo term such that the satellite fraction in these LAE samples are negligible.}
	\label{fig:acf_combined}
\end{figure*}

Shown in Figure \ref{fig:acf_IB} and \ref{fig:acf_combined} are the angular correlation functions of our intermediate band and combined intermediate band samples. We show the median angular correlation functions based on our 2000 iterations in measuring \wtheta in each figure and also include the best-fit model as described in Equation \ref{eqn:scf}.

\bsp	
\label{lastpage}
\end{document}